\documentclass[final,5p,times,twocolumn]{elsarticle}

\usepackage{amssymb}

\usepackage{textcomp}
\usepackage{multirow}
\usepackage{array}
\usepackage{graphicx}
\usepackage[caption=false]{subfig}
\usepackage{enumitem}
\usepackage{url}
\usepackage{mdwlist} 
\usepackage[resetlabels]{multibib}
\newcites{S}{Selected primary studies}
\newcites{X}{Typical excluded primary studies}

\journal{Journal of Systems and Software}

\begin{document}

\begin{frontmatter}

\title{Spot Pricing in the Cloud Ecosystem: A Comparative Investigation}


\author[zli]{Zheng Li}
\author[hzh]{He Zhang\corref{cor}}
\ead{hezhang@nju.edu.cn}
\author[liam]{Liam O'Brien}
\author[hzh]{Shu Jiang}
\author[hzh]{You Zhou}
\author[zli]{Maria Kihl}
\author[rjv]{Rajiv Ranjan}
\address[zli]{Department of Electrical and Information Technology, Lund University, Lund, Sweden}
\address[hzh]{State Key Laboratory of Novel Software Technology, Software Institute, Nanjing University, Jiangsu, China}
\address[liam]{Geoscience Australia, Canberra, ACT, Australia}
\address[rjv]{School of Computing Science, Newcastle University, Newcastle, United Kingdom}

\begin{abstract}
\textit{Background:} Spot pricing is considered as a significant supplement for building a full-fledged market economy for the Cloud ecosystem. However, it seems that both providers and consumers are still hesitating to enter the Cloud spot market. The relevant academic community also has conflicting opinions about Cloud spot pricing in terms of revenue generation.

\noindent
\textit{Aim:} This work aims to systematically identify, assess, synthesize and report the published evidence in favor of or against spot-price scheme compared with fixed-price scheme of Cloud computing, so as to help relieve the aforementioned conflict.

\noindent
\textit{Method:} We employed the systematic literature review (SLR) method to collect and investigate the empirical studies of Cloud spot pricing indexed by major electronic libraries.

\noindent
\textit{Results:} This SLR identified 61 primary studies that either delivered discussions or conducted experiments to perform comparison between spot pricing and fixed pricing in the Cloud domain. The reported benefits and limitations were summarized to facilitate cost-benefit analysis of being a Cloud spot pricing player, while four types of theories were distinguished to help both researchers and practitioners better understand the Cloud spot market.

\noindent
\textit{Conclusions:} This SLR shows that the academic community strongly advocates the emerging Cloud spot market. Although there is still a lack of practical and easily deployable market-driven mechanisms, the overall findings of our work indicate that spot pricing plays a promising role in the sustainability of Cloud resource exploitation.
\end{abstract}


\end{frontmatter}


\section{Introduction}

Cloud computing has been increasingly acknowledged in industry not only for benefiting Cloud providers by creating more business opportunities, but also for relieving Cloud consumers of purchasing, installing, and maintaining local compute resources. To guarantee a successful and sustainable Cloud ecosystem, suitable pricing techniques must be developed and implemented \cite{Weinhardt_Anandasivam_2009}. When it comes to trading Cloud resources, fixed pricing is the dominant strategy in the Cloud market nowadays \citeS{Xu_Li_2013}, \cite{Al-Roomi_2013}.\footnote{We use two types of bibliography formats: the alphabetic format denotes the Cloud service evaluation studies (primary studies) of the SLR, while the numeric format (present in the ``References" section) refers to the other references for this article.} In particular, the most common pricing scheme (namely \textit{pay as you go}) is for on-demand Cloud services, where employing a unit of service is charged a fixed price per unit of time \citeS{Abhishek_Kash_2012}. Given the normally unpredictable and stochastic demand, however, there would always be unused resources in the virtually infinite compute capacity of the Cloud. To help further and better utilize the idle compute resources, a promising approach is to provide spot resources at a reduced price so as to attract more demands with toleration of service delay and interruptions \citeS{Wang_Qi_2013}. In fact, a commercial spot market has been established when a spot instance service was launched by Amazon in December 2009 \citeS{Song_Yao_2013}. Given the de facto spot price traces that are generally far below the on-demand prices, spot pricing is claimed to be the most cost-effective scheme among the existing options for Cloud consumers. More importantly, Amazon's offering of spot service has been regarded as the first step toward a full-fledged market economy for Cloud computing \citeS{Abhishek_Kash_2012}.

Unfortunately, there seems to be a lack of confidence in becoming Cloud spot pricing players in industry. Both providers and consumers are still hesitating to enter the Cloud spot market. For instance, the overwhelming majority of the existing Cloud providers have not employed the spot pricing scheme yet \citeS{Zaman_Grosu_2011}, and the only currently available provider Amazon is still using contests to encourage more spot applications \cite{Amazon_2014}. The possible reasons for not joining the Cloud spot market could exist behind the limitations of spot pricing. Unlike the static and straightforward pricing schemes of on-demand and reserved Cloud services, the market-driven mechanism for pricing spot service would be complicated for both implementation and understanding. Moreover, since the overall supply and demand of spot resources are both uncertain during runtime, spot service consumers would have to suffer from the irregular fluctuations in service price and availability.

Meanwhile, there are also conflicting opinions in the relevant academic community. As mentioned previously, offering spot resources has been viewed as an effective approach to attracting more consumers, fully utilizing the Cloud resources, and generating more revenue \citeS{Wang_Qi_2013}. Nevertheless, some theoretical analysis and simulation argued that directly using fixed prices would bring higher expected revenues for providers than employing a hybrid (fixed + spot) pricing scheme \citeS{Abhishek_Kash_2012}. It is difficult to give a quick judgment on this even if referring to Amazon's spot service as a concrete example, because the public can obtain little information except the short-term history of spot prices. 

To help alleviate the conflict in such a background and understand whether or not it is reasonable to employ spot pricing for Cloud computing, we conducted a systematic literature review (SLR) \cite{Kitchenham_Charters_2007} in order to rigorously identify, assess, and synthesize empirical evidence in favor of or against Cloud spot pricing. In addition to analyzing the benefits and limitations of spot pricing, we also distinguished between different theories \cite{Gregor_2006} proposed to describe/predict prices or explain/prescribe pricing mechanisms in the Cloud spot market. Furthermore, we particularly investigated the fault-tolerance techniques developed to address the limitations of Cloud spot pricing.

Accordingly, the contributions of this work are mainly threefold. Firstly, the systematically summarized discussions and empirical evidence can help both Cloud providers and consumers gain a quick impression of the pros and cons of the spot pricing scheme. Moreover, this report is further able to act as a checklist to facilitate cost-benefit analysis of offering/employing spot services. 
Secondly, the four types of relevant theories can help both researchers and practitioners better understand the Cloud spot market. In practice, the price prediction techniques involved in the predictive theories would be particularly valuable for Cloud consumers to make proper biddings, while the various prescriptive theories would be able to inspire Cloud providers to develop/improve their spot pricing mechanisms. In academia, researchers may refer to the existing theories to cross check and review new studies on Cloud spot pricing. 
Thirdly, the collected fault-tolerance techniques aim to give Cloud consumers an overview about how to achieve tradeoffs between economic benefits and service availability.

The remainder of this paper is organized as follows. Section \ref{sec:related_work} briefly introduces the related work on Cloud spot pricing. Section \ref{sec:review_method} specifies the SLR method and logistics in our study. Section \ref{sec:overview} reports an overview of the reviewed studies, while Section \ref{sec:results} presents our main findings from this SLR by answering the predefined research questions. A set of possible threats to the validity of this study are highlighted in Section \ref{sec:validity_threats}. Conclusions and some future work are discussed in Section \ref{sec:conclusion}.

\section{Related work}
\label{sec:related_work}
It has been recognized that adequate pricing techniques would play a key role in the success of Cloud computing in practice \cite{Weinhardt_Anandasivam_2009}. In the de facto Cloud market, different providers have employed different strategies to attract consumers and sell their Cloud services. In general, there are three typical pricing schemes mainly based on Amazon's specification \cite{Amazon_2015}, as listed below. 
\begin{itemize}
    \item	\textbf{On-Demand Service Pricing scheme}: Cloud consumers pay a fixed cost per service unit on an hourly basis without upfront fee and long-term commitment. An analogy of this pricing scheme can be paying per view from a video on demand (VOD) service through the Internet. 
    \item	\textbf{Reserved Service Pricing scheme}: Cloud consumers pay an upfront fixed fee to ensure discounted hourly pricing for a long-term commitment of service availability (e.g., 1 year, 3 years). An analogy of this pricing scheme can be signing a two-year subscription of mobile service to receive cheaper data plans with a free phone. 
    \item	\textbf{Spot Service Pricing scheme}: Cloud consumers bid on spare resources and employ them whenever the bid exceeds the current spot price, while the employed service will be interrupted when the spot price exceeds the current bid. An analogy of this pricing scheme can be the dynamic pricing in the electricity distribution industry. 
\end{itemize}

Although the fixed pricing schemes are dominant approaches to trading Cloud resources nowadays \cite{Weinhardt_Anandasivam_2009}, spot pricing has been broadly agreed as a significant supplement for building a full-fledged market economy for the Cloud ecosystem \citeS{Abhishek_Kash_2012}. In fact, a wide consensus on efficient management of resources in our society is to avoid fixed pricing \cite{Meir_Rosenschein_2013}. As such, spot pricing (also dynamic pricing) methods have been increasingly developed and adopted in various industries ranging from airlines to electric utilities \cite{Desiraju_Shugan_1999,Elmaghraby_Keskinocak_2003}. Nevertheless, considering the unique characteristics of Cloud computing like location independence, resource virtualization and rapid elasticity, it could be improper to directly confirm the benefits of Cloud spot pricing by analogy with that in other industries. 

As a matter of fact, the recent intensive investigations into Cloud spot pricing (e.g., \citeS{Javadi_Thulasiram_2013,Leslie_Lee_2013,Zaman_Grosu_2013}) have delivered diverse and even contradictory statements. In particular, the simulation in the study \citeS{Abhishek_Kash_2012} indicates that spot pricing scheme would generate lower expected revenue for Cloud providers, which conflicts with the aforementioned consensus in most cases. Several survey studies tried to give an overview of, and comparison between different Cloud pricing schemes \cite{Al-Roomi_2013,Samimi_Patel_2011}\citeX{Karunakaran_Krishnaswamy_2015,Silva_Neto_2012}. Unfortunately, none of them emphasized empirical evidence for the comparison between spot pricing and fixed pricing. Furthermore, these survey studies delivered confusing terminology and classifications to readers. For example, various theoretical and mathematical models were directly treated as different Cloud pricing schemes, whereas those models are essentially for explaining or implementing the spot pricing scheme.

Our work focuses on the empirical evidence of spot pricing in the Cloud industry. In addition to outlining an overview of the benefits and limitations of Cloud spot pricing, we also report a theory-based classification of the relevant studies for researchers' and practitioners' reference.

\section{Review method}
\label{sec:review_method}
\begin{figure}[!t]
\centering
\includegraphics{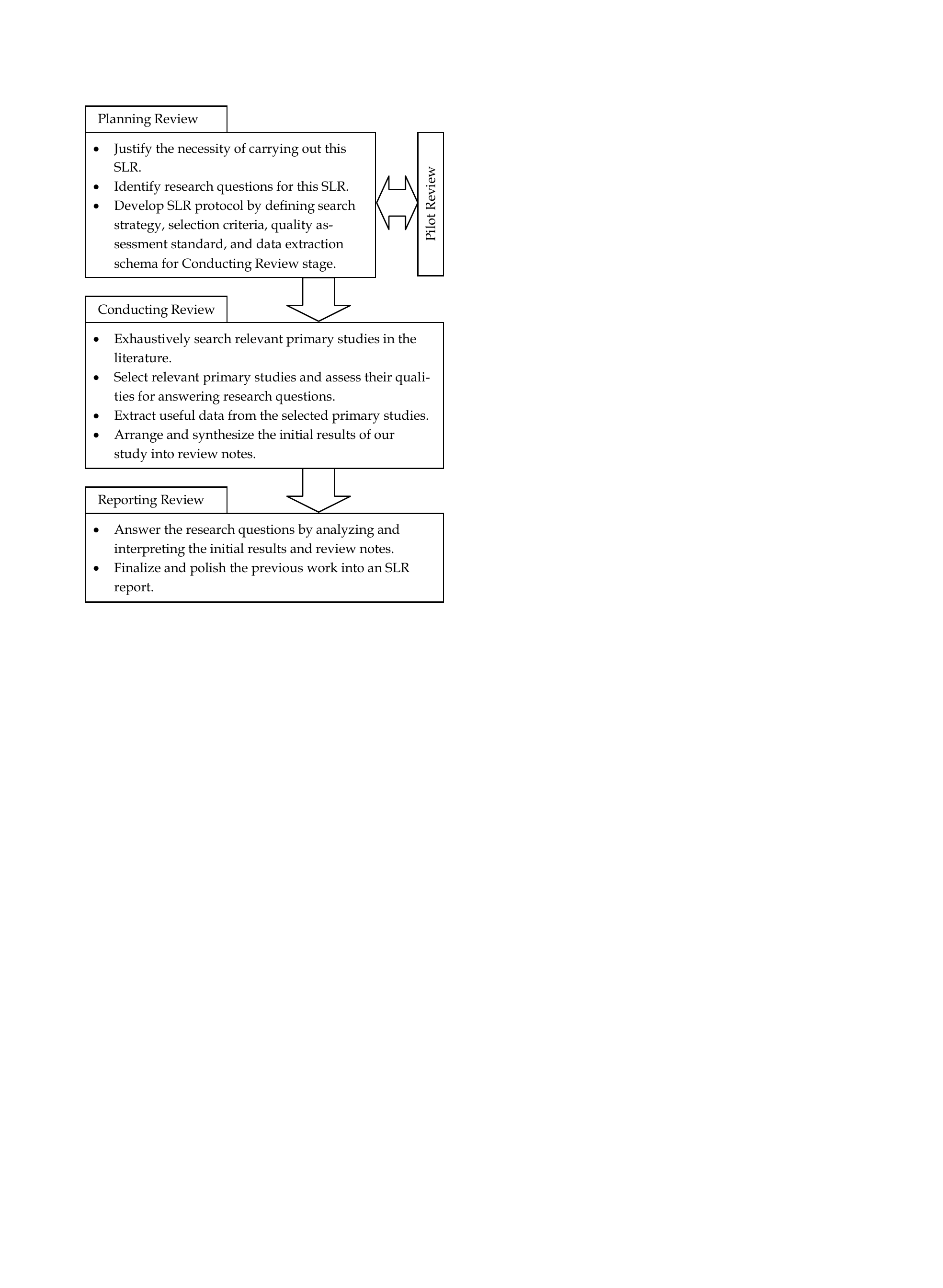}
\caption{Review procedure of this SLR.}
\label{fig_procedure}
\end{figure}

The comprehensive guidelines for performing SLR have been specified by Kitchenham and Charters \cite{Kitchenham_Charters_2007}. Here we adapted the guidelines to this work and followed a three-stage review procedure, as illustrated in Fig.~\ref{fig_procedure}. 

In particular, given the existing experiences of SLR \cite{Babar_Zhang_2009,Li_Zhang_2013}, we also emphasized the pilot review during Planning Review stage. In fact, initially reading some relevant studies would be crucial for understanding the domain knowledge and justifying the SLR work, which essentially brought the research questions. Furthermore, during the development of the review protocol, the pilot review can help gradually improve search strategy, refine inclusion/exclusion criteria, and verify data extraction schema by collecting pilot data. The remainder of this section specifies more details we prepared for conducting this SLR.

\subsection{Research questions}
\label{subsec:RQ}
There are inconsistent and even conflicting claims with regard to the benefits of applying spot pricing in the Cloud market. In order to objectively judge those claims, we decided to summarize and analyze the existing evidence, and the work was driven by the primary research question RQ1.

\begin{description}
    \item[RQ1:] \emph{What empirical evidence has been presented in the literature regarding the benefits and limitations of spot pricing in comparison to fixed pricing in the Cloud market?}
\end{description}

Moreover, as mentioned previously, the de facto mechanism for pricing spot resources in the Cloud market seems still to be a mystery. To help understand the Cloud spot market, it is worth further investigating the backend theories proposed in the relevant studies. In information systems, there always exists multiple types of theories exposing assumptions under different viewpoints, and different types of theories all can be valuable \cite{Gregor_2006}. Therefore, according to the classification of theory in \cite{Gregor_2006}, we also distinguish between four types of theories related to the Cloud spot market, as specified in Table \ref{tbl>theory}.

\begin{table*}[!t]\footnotesize
\renewcommand{\arraystretch}{1.35}
\centering
\caption{\label{tbl>theory}Four theory types related to the Cloud spot market.}
\begin{tabular}{l  >{\raggedright\arraybackslash}p{14cm}}
\hline

\hline
\textbf{Theory Type} & \textbf{Specification}\\
\hline
Descriptive Theory & This theory provides a description of the spot prices based on observations and/or statistical analyses, which essentially considers the backend spot pricing mechanism as a black box.\\

Explanatory Theory & This theory provides an explanation/clarification of the spot pricing problem, while the explanation/clarification is usually intended to promote solutions to the spot pricing problem.\\

Predictive Theory & This theory can simulate/generate spot prices without being aware of the causal demand/supply details, which essentially considers the backend spot pricing mechanism as a black box.\\

Prescriptive Theory & This theory provides an explicit prescription (e.g., methods, techniques, principles, functions, or a combination of them) for realizing the backend spot pricing mechanism.\\
\hline

\hline
\end{tabular}
\end{table*}

Given the pre-clarified theory types, we further defined the following research question to drive the corresponding investigation:

\begin{description}
    \item[RQ2:] \emph{What theories have been proposed to \mbox{describe/predict} spot prices or \mbox{explain/prescribe} spot pricing in the Cloud market?}
\end{description}

Considering that the limitations of spot pricing also imply research opportunities, there could already exist studies focusing on the potential cures. Thus, we also define a secondary research question of RQ1:  

\begin{description}
    \item[RQ3:] \emph{What techniques have been developed to address the limitations of Cloud spot pricing?}
\end{description}

Note that the research questions for this SLR were framed obeying the PICO (Population, Intervention, Comparison, Outcome) criteria \cite{Kitchenham_Charters_2007}. In detail, the population here refers to the participants in the Cloud market; the intervention is the Cloud spot pricing including its backend theories; the comparison intervention is the static strategies for pricing Cloud resources; and the outcomes that are of interest to this systematic review indicate the usefulness of spot pricing and fixed pricing within the Cloud computing domain. 

\subsection{Research scope}
\label{subsec:research_scope}
It has been identified that there are unique characteristics of and insights into Cloud computing compared to other computing paradigms \cite{Foster_Zhao_2008}. Therefore, although extensive research efforts on pricing in communication networks and Internet can be found in the literature, we only concentrate on the pricing studies in the domain of Cloud computing.

When it comes to the participants in the Cloud market, we are concerned with Cloud providers and consumers without considering their external-party connections like with power suppliers. In particular, Cloud consumers can be Cloud service brokers, secondary service providers, or end users. As for Cloud providers, recall that Amazon is the de facto vendor in the spot market at the time of writing. To make our work closer to the real situation in industry, we only consider the pricing scenario where a single Cloud provider sells spot resources to its consumers. Although a multi-provider spot market will very possibly appear in the future, we are not interested in the evidence for a concern that is not yet a problem. Moreover, different empirical evidence for different pricing scenarios could not be compatible or comparable to each other.

\subsection{Search strategy}
It has been identified that the rigor of the search strategy is crucial for its corresponding systematic reviews \cite{Zhang_Babar_Tell_2011}. A rigorous strategy can minimize the bias and increase the confidence of repeatable search process and consistent search results. To achieve this, we emphasize the following steps: setting a precise publication time span, employing popular electronic libraries, designing a search string carefully, and using a manual search to supplement the automated search. 

\subsubsection{Publication time span}
As the commercial Cloud spot market was established in December of 2009 \citeS{Song_Yao_2013}, we particularly focused on the literature published from the beginning of 2010. Given such a concrete case in industry, the collected evidence would be more valuable and convincing for their practical motivations and validations in the relevant studies. Meanwhile, considering the possible delay of publishing, we only explored studies published before mid 2015. In other words, we restricted the publication time span between \textbf{January 1st, 2010} and \textbf{June 30th, 2015}.

\subsubsection{Electronic data sources}
According to the statistics of the literature search engines \cite{Zhang_Babar_Tell_2011}, and following the referential experiences reported in the existing SLR protocols and reports, we also employed five popular electronic libraries to achieve a broad enough coverage of relevant primary studies. Note that, as described in Section \ref{subsubsec:search_method}, the selected studies found by the automated search essentially acted as secondary data sources for the subsequent manual search (i.e. reference snowballing).
\begin{itemize*}
    \item	ACM Digital Library (\url{http://dl.acm.org/})
    \item	Google Scholar (\url{http://scholar.google.com})
    \item	IEEE Xplore (\url{http://ieeexplore.ieee.org})
    \item	ScienceDirect (\url{http://www.sciencedirect.com})
    \item	SpringerLink (\url{http://www.springer.com})
\end{itemize*}

\subsubsection{Search string}
We mainly followed the approach proposed in \cite{Kitchenham_Mendes_2006} to compose the search string. Firstly, we derived major terms from the aforementioned PICO-based research questions, as listed below.

\textbf{Population:} cloud

\textbf{Intervention:} spot pricing

\textbf{Comparison:} fixed pricing

\textbf{Outcomes:} benefit, limitation, revenue, cost

Secondly, we identified the alternative spellings and synonyms for these major terms, and also supplemented the terms by checking the keywords in the pilot relevant studies. 

Thirdly, we used the Boolean OR and AND to properly integrate all the determined terms into a string for the automated search:

\textbf{cloud AND (``dynamic pricing" OR ``spot pricing" OR ``market pricing" OR ``dynamic price" OR ``spot price" OR ``market price" OR ``static pricing" OR ``fixed pricing" OR ``on-demand pricing" OR ``static price" OR ``fixed price" OR ``on-demand price" OR ``reserve price") AND (profit OR profitable OR revenue OR cost OR limit OR limitation OR benefit OR benefiting)}

\subsubsection{Search process}
\label{subsubsec:search_method}
\begin{figure}[!t]
\centering
\includegraphics{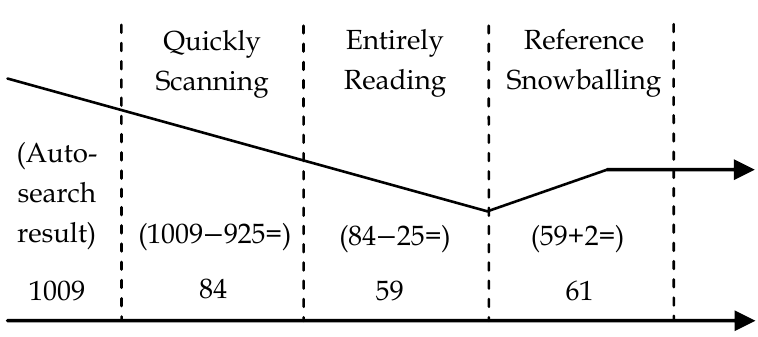}
\caption{Manual search steps (adapted from \cite{Cruzes_2011}).}
\label{fig_search_process}
\end{figure}

The search process is roughly an automated search followed by a manual search. The automated search is to run the search string through the search engines of those five popular electronic data sources. After collecting the results from automated search, we unfolded the manual search by implementing three steps, as illustrated in Fig.~\ref{fig_search_process}. Note that the manual search here essentially acts as data immersion \cite{Cruzes_2011} that facilitates data extraction of this SLR, which also confirms the importance of pilot review (cf.~Fig.~\ref{fig_procedure}). 

\begin{enumerate}
\renewcommand{\labelenumi}{\it{(\theenumi)}}
    \item	\textit{Quick Scanning:} An initial selection from the automated search results was performed by scanning their titles, keywords and abstracts against the inclusion/exclusion criteria.

    \item	\textit{Thorough Reading:} The full texts of initially selected publications were read entirely to further decide their relevance to this SLR. In particular, the unsure papers were discussed in our team meetings.

    \item	\textit{Reference Snowballing:} To explore the potential relevant studies to the largest extent, we reread the Related Work section of each selected study and then supplemented a reference snowballing \cite{Kitchenham_Li_2011}. New papers identified during reference snowballing were finally determined by repeating the previous steps.
\end{enumerate}

\subsection{Inclusion and exclusion criteria}

The inclusion and exclusion criteria were mainly shaped from the research scope (cf.~Section \ref{subsec:research_scope}) of this SLR. In detail, the criteria are specified as below:

~\\
\noindent
\textbf{\textit{Inclusion Criteria:}}
\begin{enumerate*}
\renewcommand{\labelenumi}{\it{(\theenumi)}}
    \item	Publications that investigate the de facto spot market of Cloud computing.
    \item	Publications that propose new mechanisms to support pricing/allocating/provisioning Cloud spot resources.
    \item	Publications that perform comparisons between spot pricing and fixed pricing in the Cloud market.
    \item	Publications that develop techniques to address the limitations of Cloud spot pricing.
    \item	Publications that study requesting/bidding/applying spot resources together with discussions about the Cloud spot market/mechanism/prices.
    \item	An overall condition for above inclusion criteria is that the published studies consider a single-provider pricing scenario only.
    \item	Moreover, the above inclusion criteria apply only to regular academic publications (Full journal/conference/workshop papers, technical reports, and book chapters).
\end{enumerate*}

~\\
\textbf{\textit{Exclusion Criteria:}}
\begin{enumerate*}
\renewcommand{\labelenumi}{\it{(\theenumi)}}
    \item	Publications that emphasize external connections other than Cloud provider and consumer when investigating Cloud resource pricing. To the best of our knowledge, the external connections are normally with power suppliers.
    \item	Publications that consider the competition among multiple Cloud providers, or allow one consumer to employ spot resources from multiple Cloud providers.
    \item	Publications that do not include any discussion about benefits or limitations of Cloud spot pricing.
    \item	Publications that discuss pricing for non-Cloud computing paradigms.
    \item	Publications that describe Cloud-related studies with respect to the fixed pricing only.
    \item	Publications that investigate Cloud pricing in a generic sense without distinguishing between spot and fixed mechanisms.
    \item	Publications that are previous versions of the later published work.
    \item	Survey papers (secondary studies) that do not contribute the first-hand evidence.
    \item	In addition, short/position papers, demo, posters, extended abstracts or industry publications are all excluded.
\end{enumerate*}

\subsection{Study quality assessment}
\label{subsec:quality_assessment}
Recall that it is usually difficult to distinguish between the methodological rigor and reporting quality of a research study \cite{Dyba_Dingsoyr_2008}. For the purpose of conciseness, and also by referring to the existing SLR reports, we proposed to use five assessment questions as Part I to examine a relevant study's overall quality.

~\\
\noindent
\textbf{\textit{Part I:}}

\begin{itemize*}
    \item	\textbf{\textit{QA1:}} Is there a rationale for why this study was undertaken? 
    \item	\textbf{\textit{QA2:}} Is there a description and justification for distinguishing this study from the related work?
    \item	\textbf{\textit{QA3:}} Is there a description and justification for how this study was conducted (the research design)?
    \item	\textbf{\textit{QA4:}} Is there a clear statement of the findings in this study?
    \item	\textbf{\textit{QA5:}} Does this study present sufficient data/analysis to support/justify the findings?
\end{itemize*}

Considering that this SLR work is about comparative evidence of Cloud spot pricing, we further defined three assessment questions as Part II to measure the strength of the evidence contributed by a primary study. In other words, we can use the score of Part II to reflect how relevant the primary study is to our SLR work.

~\\
\noindent
\textbf{\textit{Part II:}}
\begin{itemize*}
    \item	\textbf{\textit{QA6:}} Is there any description or justification of benefits and limitations of Cloud spot pricing?
    \item	\textbf{\textit{QA7:}} Is there any comparison between spot pricing and fixed pricing in the Cloud market?
    \item	\textbf{\textit{QA8:}} Does this study present sufficient data/analysis to support/justify the comparison?
\end{itemize*}

Answers to these quality assessment questions were assigned numerical scores 0 (`\emph{no}'), 0.5 (`\emph{to some extent}'), or 1 (`\emph{yes}'). In particular, we scored answers to the evidence strength questions 0.5 if the corresponding study only focused on Amazon, and 1 if the study discussed Cloud spot pricing in a broad sense. In addition, we noted the information locations within a paper according to which the assessments were made. As such, we easily double checked the quality scores in our meetings and resolved any disagreements.

\begin{table*}[!t]\footnotesize
\renewcommand{\arraystretch}{1.35}
\centering
\caption{\label{tbl>extraction}Data extraction schema.}

\begin{tabular}{l l >{\raggedright}p{10.3cm} >{\raggedright\arraybackslash}p{2.3cm} }
\hline

\hline
\textbf{ID} & \textbf{Data Element} & \textbf{Data Extraction Question} & \textbf{Corresponding Research Question} \\
\hline
(1) & Reference key & N/A (To help readers quickly fine a particular study.)& N/A (Metadata) \\

(2) & Publication title & What is the title of the publication?&  \\

(3) & Author & What is/are the authors' name(s)?&  \\

(4) & Affiliation & What is/are the authors' affiliation(s)? &   \\

(5) & Publication year  & In which year was the evaluation work published? & \\

(6) & Venue type  & What type of the venue does the publication have? (Journal, Conference, Workshop, Book Chapter, or Technical Report) & \\

(7) & Venue name  & Where is the publication's venue? (name of the journal, conference, workshop, or institute) & \\

(8) & Resource type & What type of Cloud spot resource is considered in this study? & N/A (Context data) \\

(9) & Demand distribution & What distribution/scenario of consumer demand is considered in this study? & \\

(10) & Application & What application is considered to consume the Cloud spot resource? & \\

(11) & Provider benefits & What are the benefits of using spot pricing for Cloud provider? & RQ1 \\

(12) & Provider limitations & What are the limitations of using spot pricing for Cloud provider?  & \\

(13) & Consumer benefits & What are the benefits of using spot pricing for Cloud consumer?& \\

(14) & Consumer limitations & What are the limitations of using spot pricing for Cloud consumer? & \\

(15) & Descriptive theory & What descriptive theory about Cloud spot pricing is proposed in this study? & RQ2 \\

(16) & Explanatory theory & What explanatory theory about Cloud spot pricing is proposed in this study? &\\

(17) & Predictive theory & What predictive theory about Cloud spot pricing is proposed in this study?  &\\

(18) & Prescriptive theory & What prescriptive theory about Cloud spot pricing is proposed in this study? &\\

(19) & Techniques & What techniques are proposed to address the limitations of Cloud spot pricing?  & RQ3 \\

(20) & Addressed limitations & What limitations of Cloud spot pricing is supposed to be addressed in this study? &\\
\hline

\hline
\end{tabular}

\end{table*}

\subsection{Data extraction}
To collect useful data to answer the previously defined research questions, we employed a data extraction schema when reviewing the primary studies, as specified in Table \ref{tbl>extraction}. We also defined a set of data extraction questions to help clarify the data element to be collected. Note that, the data extraction questions are essentially driven by the research questions of this SLR, while the context data extraction aims to help judge the strength of the identified evidence.

The data extraction work was conducted mainly by three research students under the supervision of three researchers. The extracted raw data were kept in a spreadsheet for subsequent analysis.\footnote{The extracted raw data have been shared online:\\  https://docs.google.com/spreadsheets/d/1vOcuiIHH-eNgIz0r-0b4h3qwkZgTQqV0JhqLaf1e0pw} Moreover, the data elements' locations in their corresponding publications were also marked, so as to help us quickly backtrack and validate these data collected by different reviewers. 

\subsection{Data synthesis and aggregation}
\begin{figure}[!t]
\centering
\includegraphics{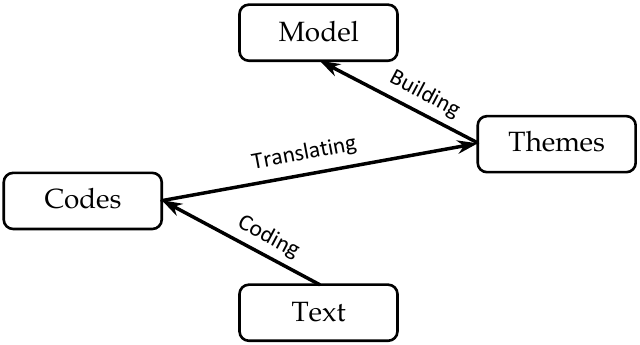}
\caption{Data evolution during thematic synthesis and aggregation.}
\label{fig_data_form}
\end{figure}

Since the raw data we collected were mostly qualitative descriptions, we employed the approach of thematic analysis \cite{Cruzes_2011} to realize data synthesis and aggregation for our SLR. As suggested by Cruzes and Dyb{\aa} \cite{Cruzes_2011}, the process of thematic synthesis and aggregation drives different forms of the data with an increasing level of abstraction, as shown in Fig.~\ref{fig_data_form}. The four types of data forms are briefly explained below.

\begin{itemize}
    \item	\textbf{Text} refers to the raw data with qualitative descriptions.
    \item	\textbf{Codes} are descriptive labels that represent different segments of the raw data.
    \item	\textbf{Themes} categorize the initial codes into a smaller set of concentrated-meaning units.
    \item	\textbf{Model} here can also denote taxonomy or theory that portrays a big picture consisting of higher-order themes and their relationships.
\end{itemize}

\section{Overview of the reviewed studies}
\label{sec:overview}
Following the search process (cf.~Section \ref{subsubsec:search_method}), we identified 61 relevant primary studies in total. In detail, the automated search initially retrieved 213, 99, 162, and 335 papers from ACM, IEEE Xplore, ScienceDirect, and SpingerLink digital libraries respectively. As for the Google Scholar, we only explored the first 20 pages of the returned results (i.e.~200 records). The manual search through Quick Skimming and Thorough Reading eventually identified 59 publications, while two more valid studies were found in the Reference Snowballing stage, as also shown in Fig.~\ref{fig_search_process}.

\begin{figure}[!t]
\centering
\includegraphics[width=7.3cm]{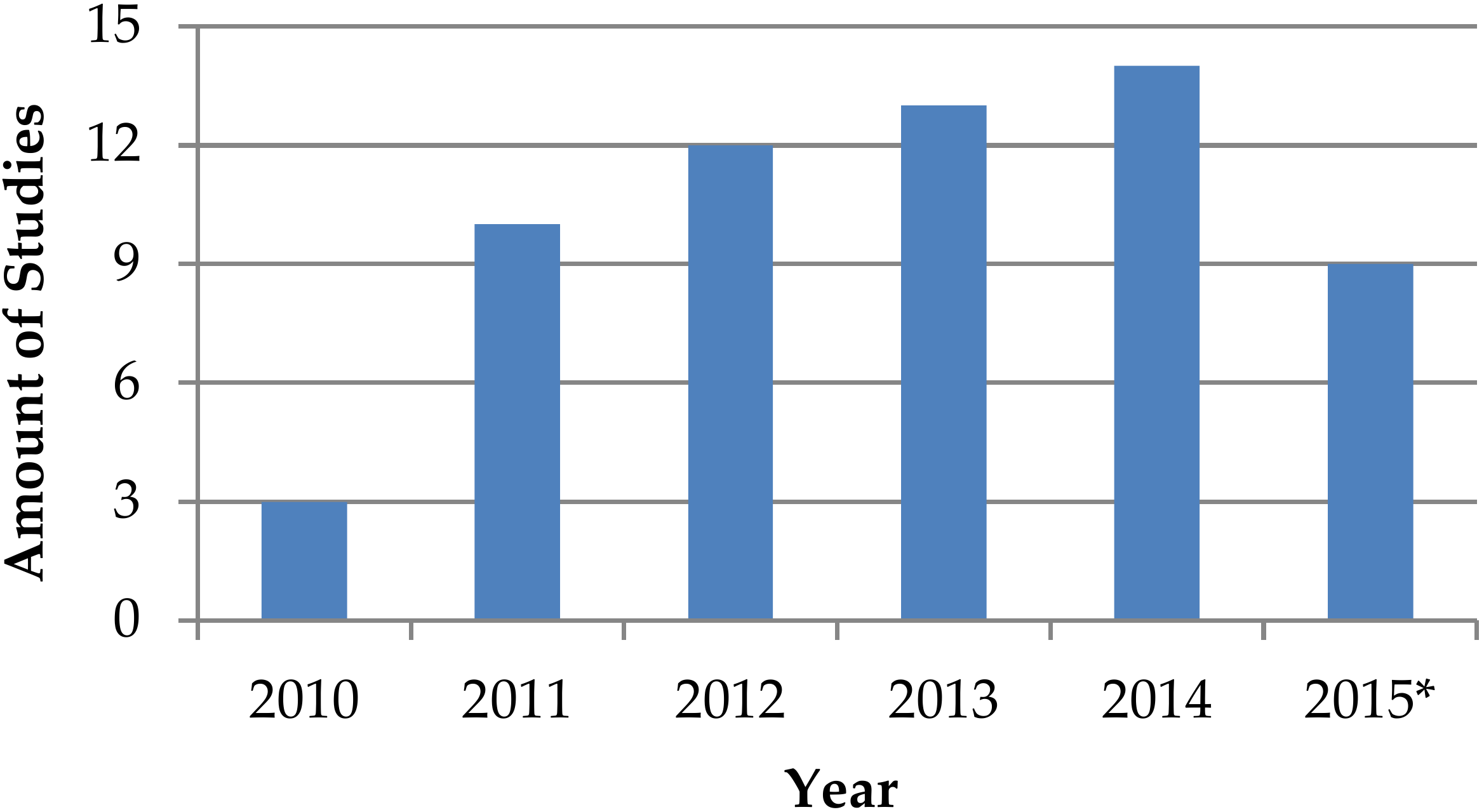}
\caption{Study distribution over the publication years. (*Note that we only explored primary studies published before mid 2015.)}
\label{fig_year_distribution}
\end{figure}

Through grouping the primary studies according to their publishing time, we show the study distribution over the past four years in Fig.~\ref{fig_year_distribution}. The rise in the number of publications indicates an increasing research interest in investigating the Cloud spot pricing. Furthermore, in spite of the various conferences related to Cloud computing, more than 20\% of the relevant studies were published in highly ranked journals, and the majority of venues (six out of eight) are IEEE/ACM transactions. Thus, it is clear that the academic community has recognized spot pricing as a significantly noteworthy research topic for satisfying the needs of Cloud industry.

\begin{figure}[!t]
\centering
\includegraphics[width=7.3cm]{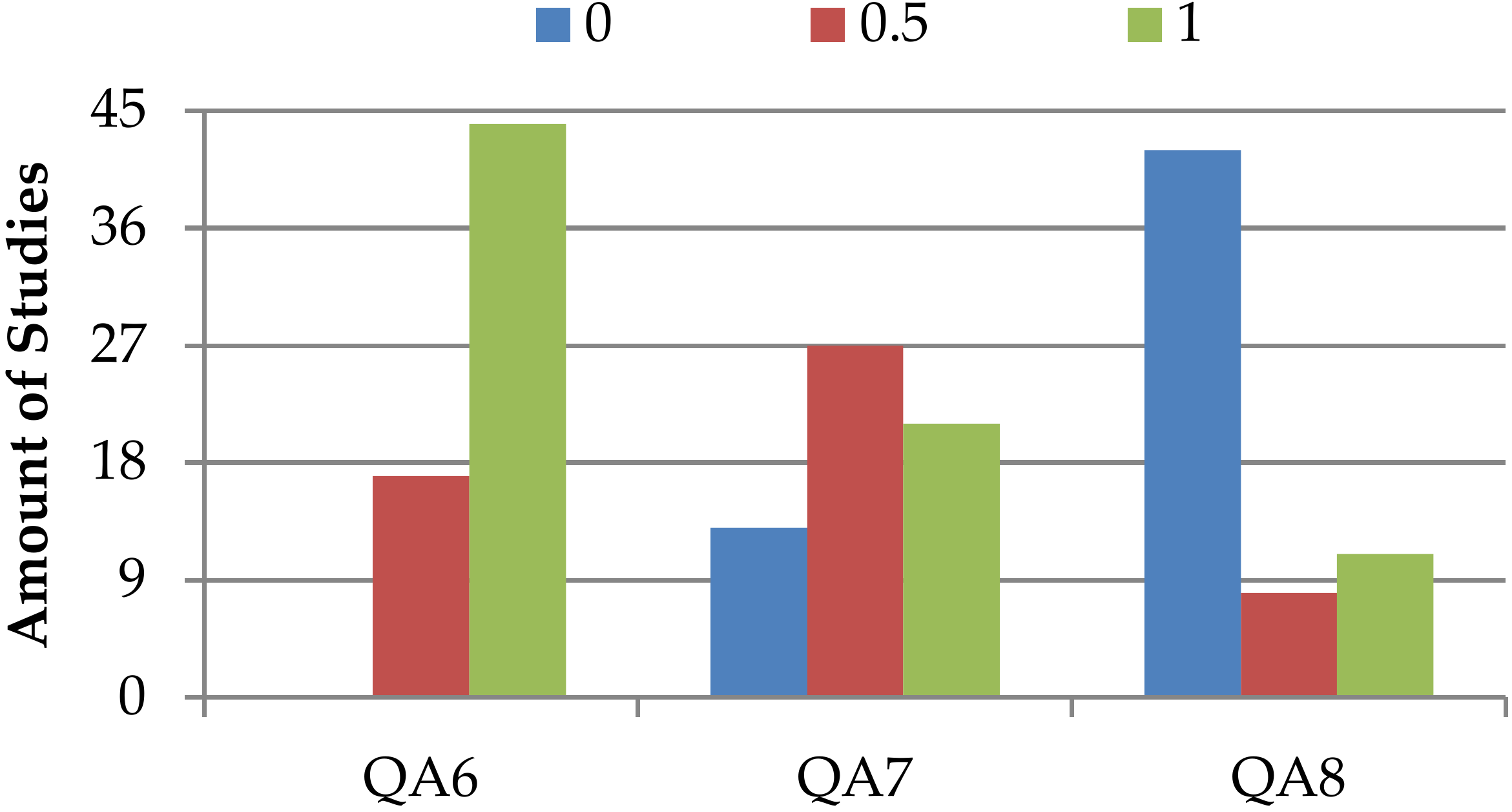}
\caption{Study distribution over the evidence strength.}
\label{fig_qa_distribution}
\end{figure}

According to the quality assessment (cf.~detailed scores in \ref{sec:quality_scores}), the selected studies were overall well conducted and reported. This confirms that our selection criteria have excluded possibly low-quality studies (e.g., short papers without sufficient evaluation). As for the strength of evidence, we further portray the study distribution over different scores in Fig.~\ref{fig_qa_distribution}. Given the \textit{Exclusion Criterion (3)}, all the relevant studies have discussed the benefits or limitations of Cloud spot pricing. However, not all of the studies performed comparison between spot and fixed pricing in the Cloud market. In particular, the scores also reveal that Amazon has naturally acted as the focus of and a concrete sample in many research efforts. More than half of the discussions and comparisons directly aimed at the de facto spot market. Only 11 studies further used simulation or statistical analysis to justify their comparisons, and most of them directly employed Amazon's spot price traces.

\section{Review results and discussions}
\label{sec:results}
The results and discussions of this SLR are organized following the sequence of answers to the three research questions. Corresponding to the data evolution process (cf.~Fig.~\ref{fig_data_form}), we categorize the benefits and limitations of spot pricing for Cloud providers and consumers respectively in Section \ref{subsec:RQ1results}, summarize four different theories for understanding the Cloud spot market in Section \ref{subsec:RQ2results}, and list currently developed techniques for addressing the limitations of Cloud spot pricing in Section \ref{subsec:RQ3results}.

\subsection{Benefits and limitations of spot pricing (RQ1)}
\label{subsec:RQ1results}
\subsubsection{Benefits for Cloud providers}
\begin{figure}[!t]
\centering
\includegraphics[width=9cm]{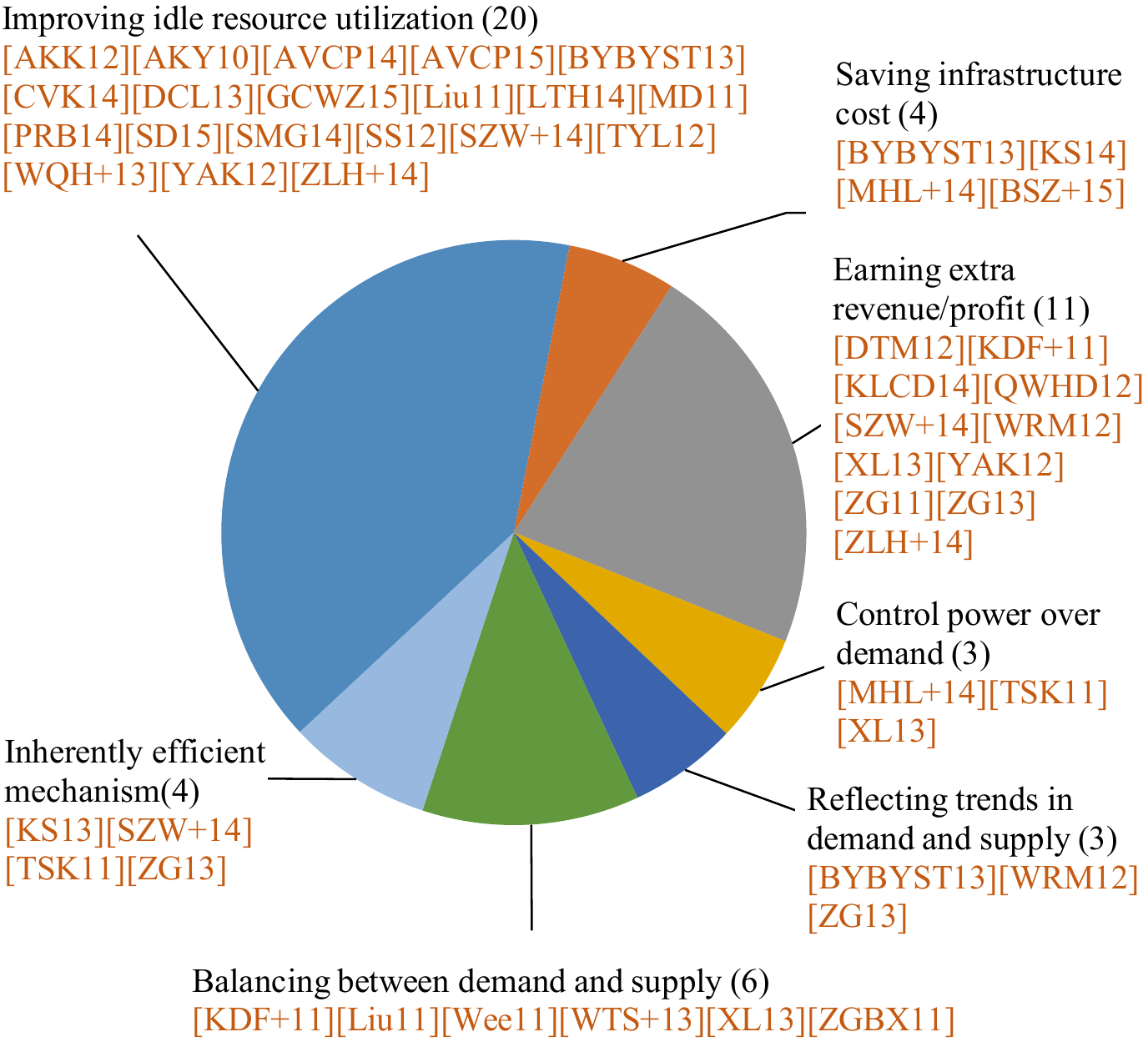}
\caption{The benefits of employing spot pricing for Cloud providers. Note that one primary study may have specified more than one benefits.}
\label{fig_providerBenefits}
\end{figure}

There are 34 out of the 61 primary studies recognizing the benefits of spot pricing for Cloud providers. We summarize the identified benefits into seven categories, as shown in Fig.~\ref{fig_providerBenefits}. 

Driven by the economics rules, dynamic pricing has been considered to be inherently more efficient than static pricing for allocating resources \citeS{Karunakaran_Sundarraj_2013}. The efficiency of resource usage would be automatically maximized under auction-based provisioning mechanisms \citeS{Taifi_Shi_2011}, because the Cloud resources can be matched to the consumers who have the highest valuation \citeS{Zaman_Grosu_2013}. In other words, the spot pricing scheme can help effectively discover the market value of Cloud resources \citeS{Shi_Zhang_2014}, especially when allocating relatively limited resources to a potentially large number of demands.

In practice, the efficiency of spot pricing for Cloud providers is mainly embodied in resource utilization and profit maximization. When it comes to the resource utilization, a clear consensus is that the spot pricing scheme can improve data centers' utilization by attracting more consumption, because it allows Cloud providers to sell spare compute capacity at lower prices which would otherwise be idle \citeS{Abhishek_Kash_2012,Andrzejak_Kondo_2010,Abundo_Valerio_2014,Abundo_Valerio_2015,Ben-Yehuda_2013,Chichin_Vo_2014,Di_Valerio_2013,Guo_Chen_2015,Karunakaran_Sundarraj_2013,Liu_2011,Lim_Thakur_2014,Mazzucco_Dumas_2011,Singh_Dutta_2015,Sadashiv_Kumar_2014,Shi_Zhang_2014,Taifi_Shi_2011,Wang_Qi_2013,Yi_Andrzejak_2012,Zhang_Li_2014}. More importantly, since fixed pricing cannot effectively reflect the underlying trends in the under- and over-demand scenarios \citeS{Wang_Ren_2012,Wallace_Turchenko_2013,Zaman_Grosu_2013}, spot pricing has been recognized as being a promising alternative to better cope with unpredictable demands and unbalance problem in the Cloud market \citeS{Xu_Li_2013,Zhang_Gurses_2011}. Through demand shifting, spot pricing can smooth out some of the computation requests with monetary incentives and lead to a more efficient use of Cloud infrastructure \citeS{Kantere_Dash_2011,Liu_2011,Wee_2011}.

As for the profit maximization, the efficiency may be further distinguished between cost saving and revenue earning. By increasing the usage of spare resources, on the one hand, spot pricing can help reduce the costs associated with idle infrastructure \citeS{Ben-Yehuda_2013,Kushwaha_Simmhan_2014,Marathe_Harris_2014}; on the other hand, spot pricing can help accommodate various consumers to generate higher revenues \citeS{Shi_Zhang_2014,Xu_Li_2013,Zaman_Grosu_2011}. Moreover, it was believed that spot resources essentially provide elasticity to the fixed-price resources without real harm to Cloud providers' main offering \citeS{Ben-Yehuda_2013}. Although Abhishek et al.~\citeS{Abhishek_Kash_2012} argued that fixed pricing is good enough in terms of revenue generation, more empirical studies have shown that under particular mechanisms the spot pricing outperforms fixed pricing by achieving more revenue for a Cloud provider \citeS{Kantere_Dash_2011,Karakus_Li_2014,Zaman_Grosu_2013,Zhang_Li_2014}, especially when allocating competitive resources to a large number of consumers \citeS{Wang_Ren_2012}.

Furthermore, spot pricing also has marginal benefits for Cloud providers. For example, the spot-price scheme gives Cloud providers a price-setting power to reclaim their compute capacity when necessary \citeS{Xu_Li_2013}, which can also be regarded as a freedom to dispatch Cloud resources to dynamic demands \citeS{Taifi_Shi_2011}. In particular, this control power prevents Cloud users from monopolizing resources through the spot market \citeS{Marathe_Harris_2014}. In some circumstances, as a marketing strategy, Cloud providers may employ imitative spot pricing to create an impression of constant change in a booming spot market, while intentionally masking the truth of low demand and price inactivity \citeS{Ben-Yehuda_2013}.

\subsubsection{Limitations for Cloud providers}
\label{subsubsec:providerLimit}

\begin{figure}[!t]
\centering
\includegraphics[width=8.7cm]{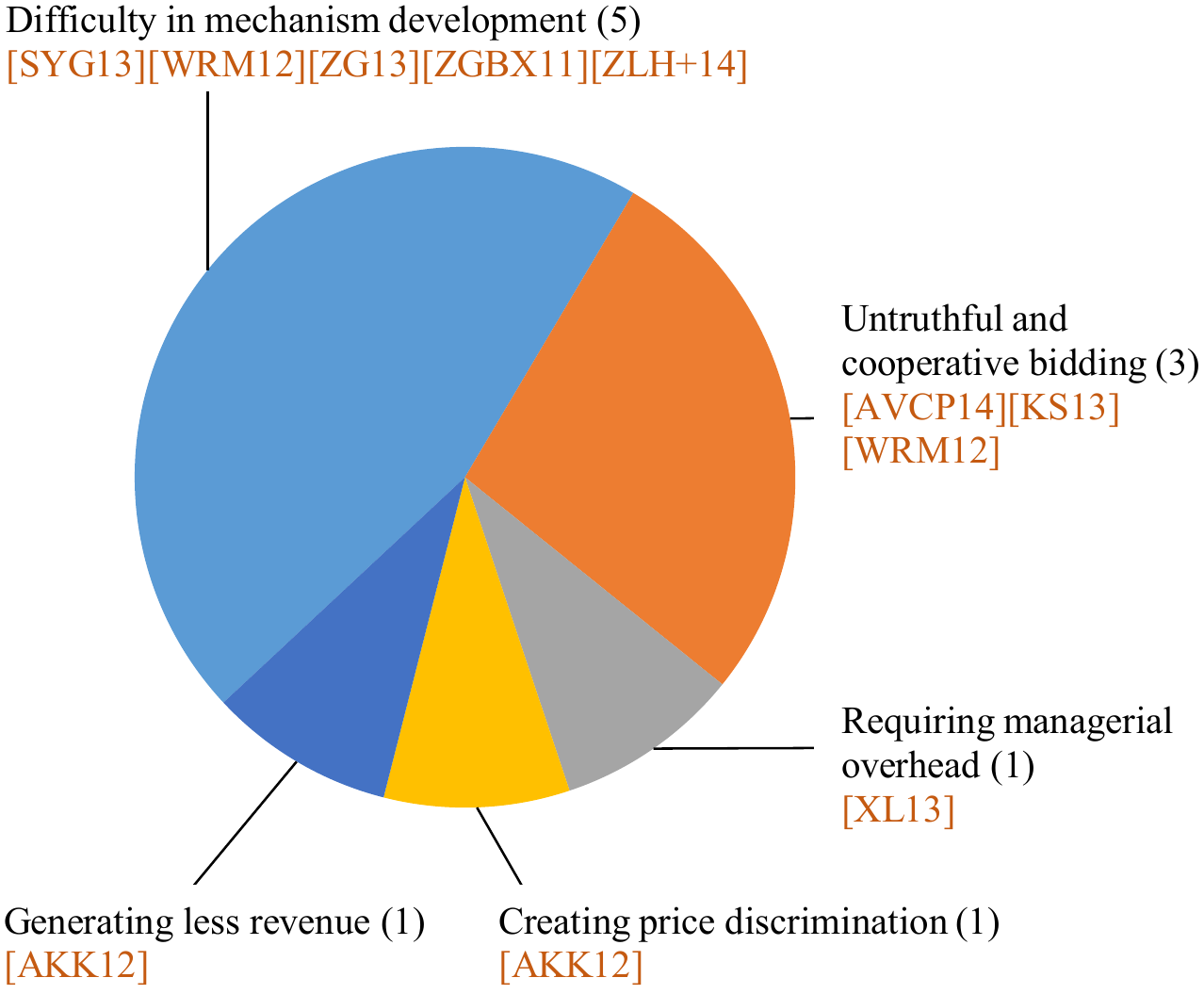}
\caption{The limitations of employing spot pricing for Cloud providers, discussed by 34 out of the 61 studies. Note that one primary study may have considered more than one limitations.}
\label{fig_providerLimits}
\end{figure}

Unlike the aforementioned benefits, the limitations of spot pricing for Cloud providers are considered by 11 studies only, and the concerns can be classified into five categories, as shown in Fig.~\ref{fig_providerLimits}.  
Compared to the straightforward fixed pricing, the spot pricing's major drawback for Cloud providers is the difficulties and challenges in developing a suitable market-driven mechanism \citeS{Zhang_Li_2014}. Suppose the spot market is formulated as auctions, no matter whether targeting revenue maximization or resource allocation, the corresponding optimization problem would be NP-hard or NP-complete \citeS{Song_Yao_2013,Wang_Ren_2012,Zaman_Grosu_2013,Zhang_Gurses_2011}. 
In addition, implementing such a market-driven mechanisms could require more effort and managerial overheads \citeS{Xu_Li_2013}. 

Furthermore, untruthful bidding and mutual cooperation can cause a cyclical effect and eventually make a market-driven strategy turn to be unprofitable for Cloud providers \citeS{Karunakaran_Sundarraj_2013,Abundo_Valerio_2014}. In fact, the de facto spot pricing in the current Cloud market has been considered to be too primitive to guarantee truthful bidding and fair resource allocation \citeS{Wang_Ren_2012}. In particular, Abhishek et al.~\citeS{Abhishek_Kash_2012} believed that operating a spot market could create price discrimination and generate less expected revenues than using fixed prices.

There is also a special concern about Amazon's policy of not charging customers for the interrupted partial hours, if the spot service usage is terminated by Amazon \cite{Amazon_2014}. In the worst case, the uncharged time rises up to 30\% of a spot instance's total run time \citeS{Dawoud_Takouna_2012}, which would become a burden and reduce the provider's profit \citeS{Sadashiv_Kumar_2014}. However, since this is Amazon's own policy, we do not treat this concern as a limitation of spot pricing in a generic sense.

\subsubsection{Benefits for Cloud consumers}

\begin{figure}[!t]
\centering
\includegraphics[width=9cm]{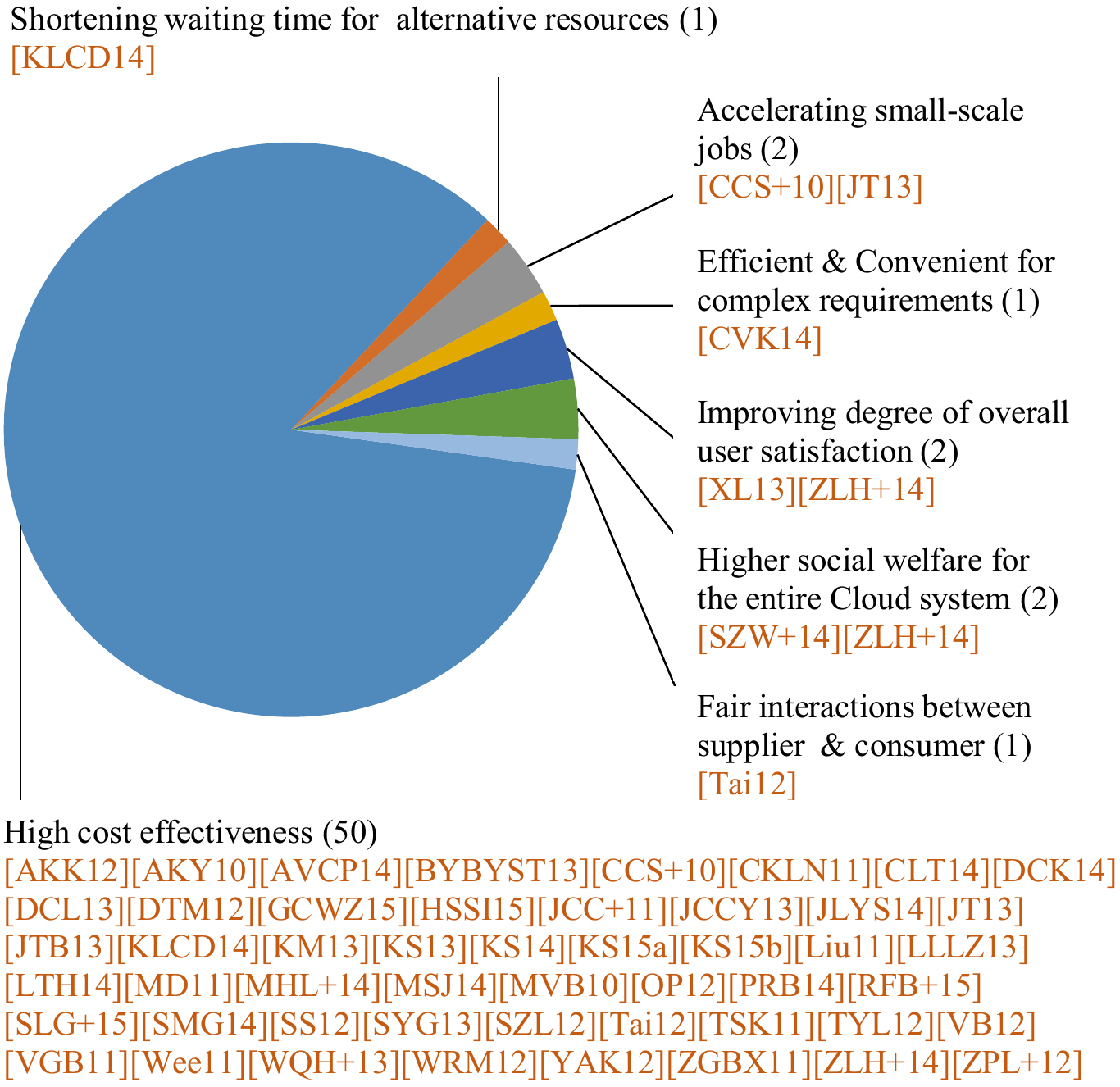}
\caption{The benefits of employing spot pricing for Cloud consumers, discussed by 53 of the 61 studies. Note that one primary study may have specified more than one benefits.}
\label{fig_consumerBenefits}
\end{figure}

As shown in Fig.~\ref{fig_consumerBenefits}, the benefits for Cloud consumers were discussed by 53 out of the 61 studies, and most of the discussions concentrated on the cost effectiveness of using spot services. Although the de facto spot service of Amazon may still unexpectedly charge more than the on-demand option \citeS{Chaisiri_Kaewpuang_2011}, it has been widely accepted that spot resources are on average cheaper (e.g., \citeS{Jung_Chin_2011,Kaminski_Szufel_2015,Mazzucco_Dumas_2011,Poola_Ramamohanarao_2014,Song_Yao_2013}). Such a consensus is mainly based on the investigation into the current Cloud spot market. For example, the quantitative analyses of Amazon's price history show that consumers can expect to save more than half the expense if replacing on-demand instances with the spot ones \citeS{Ben-Yehuda_2013,Javadi_Thulasiram_2013,Song_Yao_2013}. The empirical studies deliver even more encouraging results: with proper bids, the total cost of employing spot resources can be maintained between 13\% and 36\% of using the equivalent on-demand resources \citeS{Leslie_Lee_2013,Mattess_Vecchiola_2010}. 

Meanwhile, the spot pricing scheme has been claimed to be able to improve customer satisfaction on overall Cloud performance \citeS{Xu_Li_2013,Zhang_Li_2014}. Firstly, if there is a shortage of on-demand resource, Cloud users can shorten their waiting time by finding alternative resources in the spot market \citeS{Karakus_Li_2014}. Secondly, spot instances have been considered as accelerators for small-scale jobs. For example, by employing suitable fault-tolerant technique, the study \citeS{Jangjaimon_Tzeng_2013} shows that the application turnaround time can be shortened by up to 58\% and lowering the monetary cost by up to 59\%; by employing additional spot nodes in the MapReduce process, the study \citeS{Chohan_Castillo_2010} shows that the speedup for the overall MapReduce time of some workloads can exceed 200\% with an extra monetary cost of 42\% \citeS{Chohan_Castillo_2010}. Thirdly, if a spot market implements the mechanism of combinatorial auction, trading different types of Cloud resources would be more efficient and more convenient for the consumers with complex requirements \citeS{Chichin_Vo_2014}.

Moreover, the free market structure behind spot pricing can ensure fair interactions between Cloud providers and consumers \citeS{Taifi_2012}. The market-driven dynamic resource provisioning also enables higher social welfare for the entire Cloud ecosystem \citeS{Shi_Zhang_2014,Zhang_Li_2014}.  

\subsubsection{Limitations for Cloud consumers}
\label{subsubsec:consumerlimit}

\begin{figure}[!t]
\centering
\includegraphics[width=8.9cm]{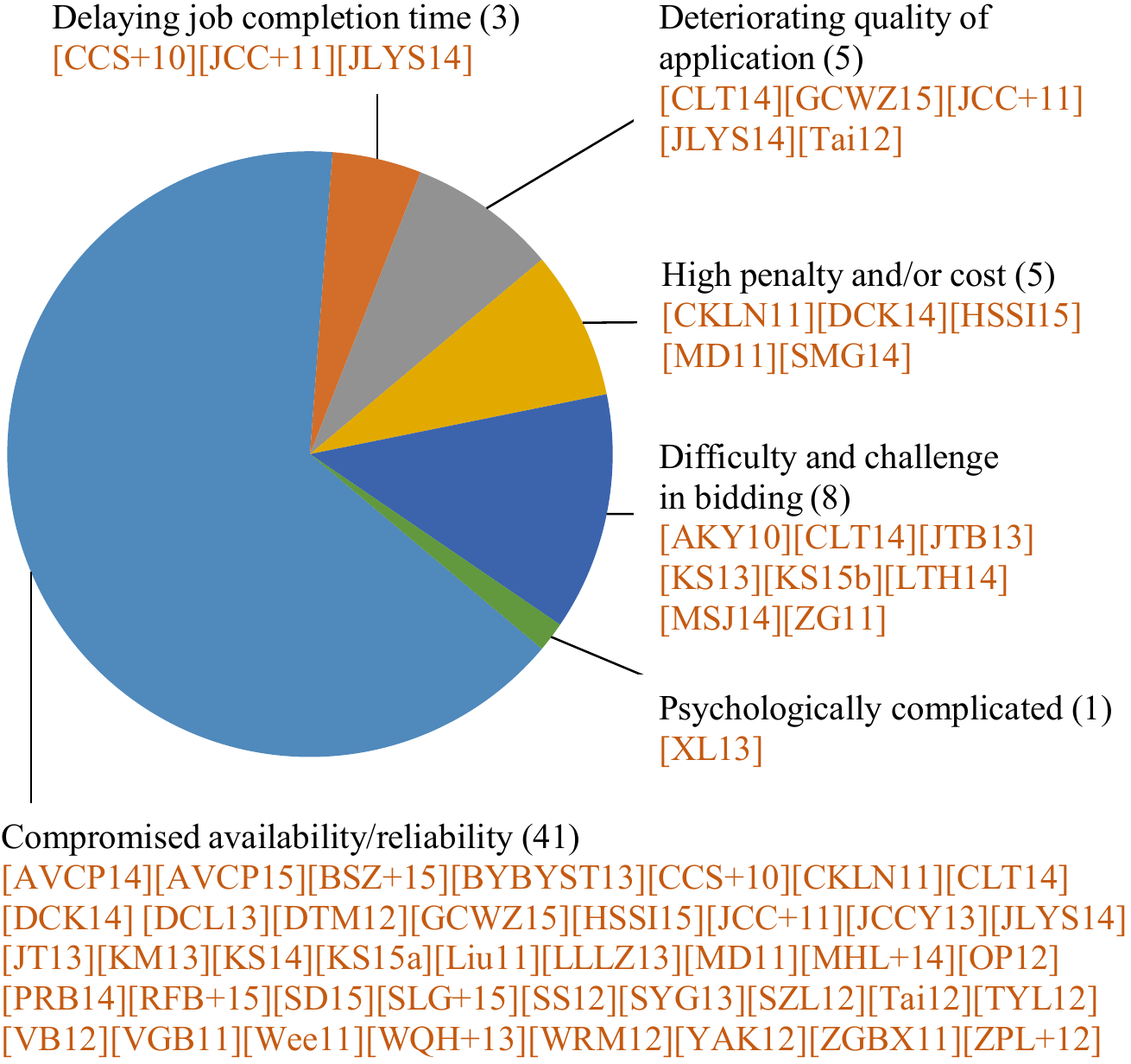}
\caption{The limitations of employing spot pricing for Cloud consumers, discussed by 50 out of the 61 studies. Note that one primary study may have considered more than one limitations.}
\label{fig_consumerLimits}
\end{figure}

Similar to the benefits for Cloud consumers, the limitations of spot pricing were highlighted also by most of the selected studies (50 out of 61), as distributed in Fig.~\ref{fig_consumerLimits}. The most significant limitation consumers have to face is the compromised availability/reliability of Cloud spot services. Given the nature of spot pricing \cite{Amazon_2014}, a Cloud provider has the control over terminating spot services through price adjustment. In other words, a spot service can become unavailable at any time without any notice due to the supply and demand fluctuations \citeS{Abundo_Valerio_2014,Abundo_Valerio_2015,Chaisiri_Kaewpuang_2011,Di_Valerio_2013,Khatua_Mukherjee_2013_paper,Voorsluys_Buyya_2012,Voorsluys_Garg_2011,Wang_Ren_2012,Yi_Andrzejak_2012,Zhang_Gurses_2011}. In fact, Amazon's spot price history indicates that consumers may still experience out-of-bid events even if their bid is as high as the regular prices \citeS{Chaisiri_Kaewpuang_2011,Javadi_Thulasiram_2013,Lim_Thakur_2014}. 

Consequently, it will become hard to analyze the availability level of services or applications built on the top of spot instances \citeS{Guo_Chen_2015}, and the completion time of job processing could be seriously delayed \citeS{Jung_Chin_2011,Chohan_Castillo_2010,Jung_Lim_2014}. From the end user's point of view, the quality of those spot-instance-based services/applications would have been deteriorated due to their uncertain performance, which inevitably leads to negative impacts on the economic advantages of using spot service for two reasons: Firstly, the quality deterioration might incur penalties for failing to meet performance and availability objectives \citeS{He_Shenoy_2015,Mazzucco_Dumas_2011}. Secondly, if not used carefully, the cheaper spot resources with frequent interruptions would eventually be more expensive than fix-priced, on-demand instances with respect to the overall cost of a Cloud-based application \citeS{Chaisiri_Kaewpuang_2011,Dadashov_Cetintemel_2014,Sadashiv_Kumar_2014}.

As such, spot pricing has been considered not to be a suitable scheme for workloads with little flexibility \citeS{Xu_Li_2013}, and spot resources have been particularly reported as inappropriate for supporting long running jobs \citeS{Chaisiri_Kaewpuang_2011,Khatua_Mukherjee_2013_paper}.

Additionally, spot pricing seems complicated for Cloud consumers to understand psychologically \citeS{Xu_Li_2013}, and therefore the inherent complexity could make it difficult to carry out a wise bidding \citeS{Andrzejak_Kondo_2010,Chen_Lee_2014,Javadi_Thulasiram_2013,Karunakaran_Sundarraj_2013,Karunakaran_Sundarraj_2015,Lim_Thakur_2014,Menache_Shamir_2014,Zaman_Grosu_2011}. Indeed, Cloud consumers might always face a dilemma when making bidding decisions. On the one hand, bidding low prices can result in extremely long execution time without reducing much monetary cost \citeS{Andrzejak_Kondo_2010}. On the other hand, bidding high prices can lead to a large cost increase without decreasing much computation time \citeS{Kaminski_Szufel_2015}. In particular, bidding above on-demand price is claimed to be unuseful, although it might be intuitively attractive \citeS{Guo_Chen_2015,Karunakaran_Sundarraj_2015}.

\begin{table*}[!t]\footnotesize
\renewcommand{\arraystretch}{1.3}
\centering
\caption{\label{tbl>benefit_limitation}Benefits and limitations of Cloud spot pricing.}
\begin{tabular}{l  >{\raggedright}p{7cm} >{\raggedright\arraybackslash}p{6cm}}
\hline

\hline
\textbf{Cloud Spot Pricing} & \textbf{Benefits} & \textbf{Limitations}\\
\hline
\textbf{For Providers} & 
\begin{itemize}[leftmargin=*]
    \item	Efficient resource utilization
    \begin{itemize*}
    		\item	Inherently efficient market-driven mechanism
    		\item	Improving utilization of idle infrastructure
    		\item	Reflecting trends in changing demand and supply
    		\item	Balancing between changing demand and supply
    \end{itemize*}
    \item	Profit Maximization
    \begin{itemize*}
    		\item	Saving cost associated with idle infrastructure
    		\item	Earning revenue by attracting more consumers
    \end{itemize*}
    \item	Control power over unpredictable demand
\end{itemize} & 
\begin{itemize}[leftmargin=*]
    \item	Difficulty in developing a suitable mechanism
    \item	Leading to untruthful and cooperative bidding
    \item	Requiring managerial overhead
    \item	Creating price discriminations
    \item	Generating lower expected revenue\textsuperscript{*}
\end{itemize} \\
\hline
\textbf{For Consumers} & 
\begin{itemize}[leftmargin=*]
    \item	High cost effectiveness
    \item	Improving customer satisfaction
    \begin{itemize*}
    		\item	Shortening waiting time for alternative resources
    		\item	Accelerating small-scale jobs
    		\item	Efficient \& Convenient for complex requirements
    		\item	improving degree of overall user satisfaction 
    \end{itemize*}
    \item	Higher social welfare for the entire Cloud system
    \item	Fair interactions between supplier \& consumer
\end{itemize} & 
\begin{itemize}[leftmargin=*]
    \item	Compromised service availability/reliability
    \begin{itemize*}
    		\item	Increasing job completion time
    		\item	Deteriorating quality of application 
    		\item	Incurring high penalty and/or cost
    \end{itemize*}
    \item	Difficulty in making bidding wisely 
    \item	Psychologically complicated pricing logic
\end{itemize} \\
\hline

\hline
\multicolumn{3}{l}{\textsuperscript{*}\footnotesize{Only one study claims that fixed-price scheme can generate higher expected revenue \citeS{Abhishek_Kash_2012}.}}\\
\end{tabular}
\end{table*}

\begin{figure}[!t]
\centering
\includegraphics[width=7.5cm]{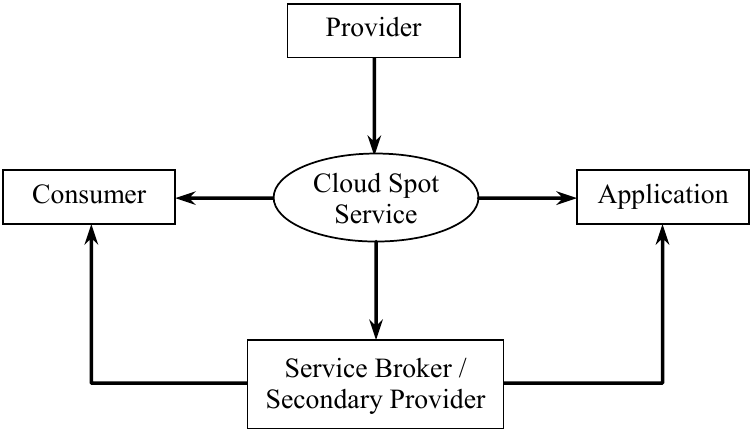}
\caption{Spot service consumption chain in the Cloud ecosystem.}
\label{fig_consumptionChain}
\end{figure}

\subsubsection{Discussion}
\label{subsubsec:RQ1_discusion}
For the convenience of reading, we concisely highlight the collected benefits and limitations of Cloud spot pricing in Table~\ref{tbl>benefit_limitation}.

Through identifying these opinions about Cloud spot pricing, we find a clear consumption chain in the Cloud ecosystem, as shown in Fig.~\ref{fig_consumptionChain}. In addition to the investigations from the provider's and consumer's perspectives, the existing studies were also concerned with concrete applications, and a service broker (or secondary provider) that acted as a middleware between spot resources and their end consumption. 

Although the relevant studies were unfolded from various aspects of the spot service consumption chain, a large degree of consensus has been reached on the benefits and limitations of Cloud spot pricing.

When it comes to the benefits of spot pricing, although different beforehand assumptions may make different studies incomparable, the majority of empirical evidence advocates that spot pricing would be more profitable. A conflicting opinion is that the fixed-price scheme can generate higher expected revenue for Cloud providers \citeS{Abhishek_Kash_2012}. Since suitable mechanisms should be a prerequisite of implementing spot pricing, it is possible that the simulation results could have been flawed by the improper assumptions implementation in \citeS{Abhishek_Kash_2012}. In fact, by analogy, the spot service pricing has been successfully applied to different industrial fields \cite{Desiraju_Shugan_1999}.

Similarly, the discussions about limitations of offering spot pricing were also intensive. A typical limitation is that the complexity in backend mechanisms could prevent both Cloud providers and consumers from joining the spot market. Furthermore, Cloud consumers are particularly concerned with unpredictably frequent interruptions when using spot services, although in some optimistic scenarios spot instances could be as reliable as standard instances.

\subsection{Theories of Cloud spot pricing (RQ2)}
\label{subsec:RQ2results}
\subsubsection{Descriptive Theories}
\label{subsubsec:descriptive}
Descriptive theories only focus on the spot prices or the variation in spot prices (cf.~Table \ref{tbl>theory}). Given the observations on Amazon's price history, a major standpoint is that there is no law behind the spot price variation \citeS{Mazzucco_Dumas_2011,Song_Zafer_2012}, and correspondingly the spot prices are totally uncertain \citeS{Chaisiri_Kaewpuang_2011} or randomly generated within a particular band \citeS{Ben-Yehuda_2013,Mattess_Vecchiola_2010}. The spot service reliability varies strongly depending on bid price, instance type, and availability zone \citeS{Andrzejak_Kondo_2010,Ben-Yehuda_2013,Lim_Thakur_2014,Ribas_Furtado_2015,Kaminski_Szufel_2015}.

To simplify their research work, some studies assumed that spot prices followed a mathematical distribution like normal distribution \citeS{Mazzucco_Dumas_2011} or standard Gaussian distribution \citeS{Menache_Shamir_2014}, and the spot price variations followed a semi-Markovian process \citeS{Song_Zafer_2012,Guo_Chen_2015}. However, several statistical analyses have revealed that the normal distribution does not fit well the price history \citeS{Zhao_Pan_2012}, while the Mixture of Gaussians (MoG) distribution delivers better price approximation compared to the other distribution models \citeS{Javadi_Thulasiram_2013}. In addition to the complex MoG distribution, the lognormal distribution can also adequately model the spot-price distribution \citeS{Karunakaran_Sundarraj_2015}.

\subsubsection{Explanatory theories}
\label{subsubsec:explanatory}
When explaining the spot pricing problem, this type of theories proposed by different studies vary significantly. 

Following the initial description about Cloud spot market \cite{Amazon_2014}, the most straightforward idea is to consider spot prices as results from auctions \citeS{Abundo_Valerio_2015,Chichin_Vo_2014,Kushwaha_Simmhan_2014,Karunakaran_Sundarraj_2015,Marathe_Harris_2014,Poola_Ramamohanarao_2014,Singh_Dutta_2015}. For example, since different consumers' bids are unknown to each other and the identical spot resources are sold at an identical price, the Cloud spot market was viewed as a continuous sealed-bid uniform price auction model \citeS{Zhang_Gurses_2011}; considering that only losers are allowed to repetitively submit new bids while winners have to remain with their positions, Song et al. \citeS{Song_Yao_2013} treated the spot market as a modified version of the repeated single-price auction; suppose the top-$N$ bidders win and the spot price equals to the highest unsuccessful bid ($(N+1)$st bid), the spot market can be modeled as a $(N+1)$st price auction with multiple goods \citeS{Ben-Yehuda_2013,Karunakaran_Sundarraj_2015}; while if taking into account the nature of diverse demands of different numbers and types of Cloud resource, the combinational auction was claimed to be best suited for representing the allocation of Cloud spot resources \citeS{Wang_Ren_2012,Zaman_Grosu_2013,Shi_Zhang_2014}.

Consequently, game theories have been further used to model the competitions that may happen within auctions. For instance, the Prisoner Dilemma game and the Generalized Nash Equilibrium (GNE) game was employed to formulate the conflicts between a provider and its consumers \citeS{Di_Valerio_2013,Karunakaran_Sundarraj_2013,Sowmya_Sundarraj_2012}; and particularly, inspired by the attacker-defender scenarios, the situation of a single Cloud provider with multiple consumers was modeled as a Stackelberg game \citeS{Di_Valerio_2013}.

The existing studies also tried to specify the Cloud spot market from engineering viewpoints. For example, Abhishek et al.~modeled the spot market as $k$ parallel $M/M/1$ queues \citeS{Abhishek_Kash_2012}, and another work regarded the market as a faulty machine driven by a semi-Markovian process with up and down states \citeS{Song_Zafer_2012}. 

From the perspective of Cloud provider, the spot pricing problem has been often translated into revenue maximization problems \citeS{Wang_Qi_2013}. In particular, the revenue maximization was further formulated as a finite-horizon stochastic dynamic program \citeS{Xu_Li_2013}; the virtual resource allocation was formed into a multi-dimensional knapsack problem (MKP) with the additional constraints \citeS{Chichin_Vo_2014}; while Kantere et al.~modeled optimal pricing as an optimal control problem with a finite horizon, and the demand curve was modeled by employing second order differential equations with constant parameters \citeS{Kantere_Dash_2011}.

Nevertheless, a reverse engineering study doubted about the market-driven mechanism behind the de facto Cloud spot market \citeS{Ben-Yehuda_2013}. Given the nearly identical trend of price changing history of different spot instance types at various regions, it was believed that the spot prices were driven by a dynamic reserve price algorithms rather than depending on the market activities claimed by Amazon. The pricing algorithm upgrade might be rolled out in one data center first, followed by other data centers \citeS{Kushwaha_Simmhan_2014}. 

\subsubsection{Predictive theories}
\label{subsubsec:predictive}

By using the empirical data from Amazon's price history, the regressive process seems to be a common technique to build predictive theories. Through reverse engineering, an auto-regressive algorithm AR(1) was developed to simulate spot price generation, and the generation results showed positive match with Amazon's price traces \citeS{Ben-Yehuda_2013}. Unlike AR(1) that tries to keep a linear relation between service availability and prices, the autocorrelation function (ACF)-based algorithm focused on the relation between price variation and the corresponding time difference, which also delivered encouraging prediction performance \citeS{Mazzucco_Dumas_2011}. On the contrary, the analysis of spot price predictability claimed that using an auto-regressive moving average model could not yield satisfactory accuracy of price approximation \citeS{Zhao_Pan_2012}. To achieve better accuracy, a Multi-Layer-Perceptron (MLP) model with a moving simulation mode was developed for short-term price prediction only \citeS{Wallace_Turchenko_2013}.   

In addition to direct spot price predictions, the predictive theory can also be spot instance failure estimations. For example, by modeling the failure probability of a spot instance based on a semi-Morkovian process of spot prices, the spot instance's failure probability for the next bidding interval can be estimated under a specific bid, which essentially implies the price prediction \citeS{Guo_Chen_2015}.

\begin{table*}[!t]\footnotesize
\renewcommand{\arraystretch}{1.3}
\centering
\caption{\label{tbl>detailedTheories}Detailed theories of Cloud spot pricing.}
\begin{tabular}{>{\raggedright}p{1.8cm}  >{\raggedright}p{6.8cm} >{\raggedright\arraybackslash}p{8cm}}
\hline

\hline
\textbf{Cloud Spot Pricing} & \textbf{About the Past} & \textbf{About the Future}\\
\hline
\textbf{Black-box Theories} & 
Descriptive theory:
\begin{itemize}
    \item	Price variations are uncertain
    \item	High price variability between zones and between instance types
    \item	Price variations follow a semi-Markovian process
    \item	Prices follow the normal distribution
    \item	Prices follow the standard Gaussian distribution
    \item	Prices follow a mixture of Gaussians distribution
    \item	Prices follow the lognormal distribution
\end{itemize} & 
Predictive Theory:
\begin{itemize}
    \item	Linear function of service availability and price
    \item	Autocorrelation function of price variation and time difference
    \item	Multi-Layer-Perceptron model with a moving simulation mode
    \item	Semi-Morkovian process-based failure estimation
\end{itemize} \\
\hline
\textbf{White-box Theories} & 
Explanatory theory:
\begin{itemize*}
    \item	Auction models
			\begin{itemize*}
    				\item	Continuous sealed-bid uniform price auction
    				\item	Repeated single-price auction
    				\item	$(N+1)$st price auction with multiple goods
    				\item	Combinational auction
			\end{itemize*}
    \item	Game theories 
			\begin{itemize*}
    				\item	Prisoner Dilemma game
    				\item	Generalized Nash Equilibrium game
    				\item	Stackelberg game
			\end{itemize*}
    \item	Mechanical models 
			\begin{itemize*}
    				\item	K parallel M/M/1 queues
    				\item	A faulty machine
			\end{itemize*}
    \item	Revenue maximization problems
			\begin{itemize*}
    				\item	Finite horizon stochastic dynamic problem
    				\item	Optimal control problem
    				\item	Multi-dimensional knapsack problem (MKP)
			\end{itemize*}
    \item	Unknown dynamic reserve price algorithm
\end{itemize*} & 
Prescriptive theory:
\begin{itemize}
    \item	Resource provisioning \& allocation algorithm (for auction models)
			\begin{itemize*}
    				\item	CA-PROVISION
    				\item	Greedy allocation + collusion-resistant algorithm
    				\item	Order-statistic-based online pricing algorithm
    				\item	Multi-round combinatorial auction framework
			\end{itemize*}
    \item	Equilibrium-related methods from Economics (for game theories)
    \item	Optimization solutions (for revenue maximization problems)
			\begin{itemize*}
    				\item	Lyapunov optimization algorithm
    				\item	Hamilton-Jacobi condition based approach
    				\item	Mixed-integer nonlinear programming
    				\item	N-armed bandit $\epsilon$-greedy approach
			\end{itemize*}
\end{itemize} \\
\hline

\hline
\end{tabular}
\end{table*}

\subsubsection{Prescriptive theories}
\label{subsubsec:prescriptive}
The prescriptive theories in this case are normally proposed solutions to the previously formulated spot pricing problems. 

Corresponding to the auction models, the provisioning and allocation of spot resources were mostly emphasized for proposing pricing mechanisms, such as CA-PROVISION \citeS{Zaman_Grosu_2013}, greedy allocation algorithm, collusion-resistant algorithm \citeS{Chichin_Vo_2014,Wang_Ren_2012}, and a multi-round combinatorial auction framework \citeS{Shi_Zhang_2014}; while Song et al.~\citeS{Song_Yao_2013} developed an order-statistic-based online pricing (OSOP) algorithm to determine spot prices without considering consumer behaviors. 

With respect to the game theories, equilibrium-related mathematical methods were directly adapted from the economics field to the Cloud case (e.g., solving variational inequality \citeS{Di_Valerio_2013}). 

When it comes to solving the revenue maximization problem, Wang et al.~\citeS{Wang_Qi_2013} adopted a Lyapunov optimization algorithm with the basic idea of minimizing a bound on the drift-plus-penalty term; Xu and Li \citeS{Xu_Li_2013} proposed an optimal solution based on a demand model, the Hamilton-Jacobi conditions, and a standard numerical approach; and Kantere et al.~\citeS{Kantere_Dash_2011} employed mixed-integer nonlinear programming to optimize the service pricing; and in one of the experimental scenarios, the provider was assumed to set spot prices using an N-armed bandit $\epsilon$-greedy approach that repeatedly select the greedy pricing action among a set of N actions \citeS{Abundo_Valerio_2014}.

Overall, the sophisticated prescriptive theories here confirm that developing a practical spot pricing mechanism would still be challenging, because both Cloud providers and consumers often prefer simplicity in practice \citeS{Kushwaha_Simmhan_2014}.

\subsubsection{Discussion}
\label{subsubsec:RQ2_discusion}

Similarly, we further highlight the identified theories of Cloud spot pricing in Table \ref{tbl>detailedTheories}. In essence, the four types of theories come from four different angles of the spot market. On the one hand, the descriptive and predictive theories are more consumer-oriented, while the explanatory and prescriptive theories are more provider-oriented. On the other hand, the descriptive and explanatory theories are developed mainly to reveal the knowledge of the existing Cloud spot prices, while the predictive and prescriptive theories are proposed mainly to aim at future spot prices. 

\begin{figure}[!t]
\centering
\includegraphics[width=7.2cm]{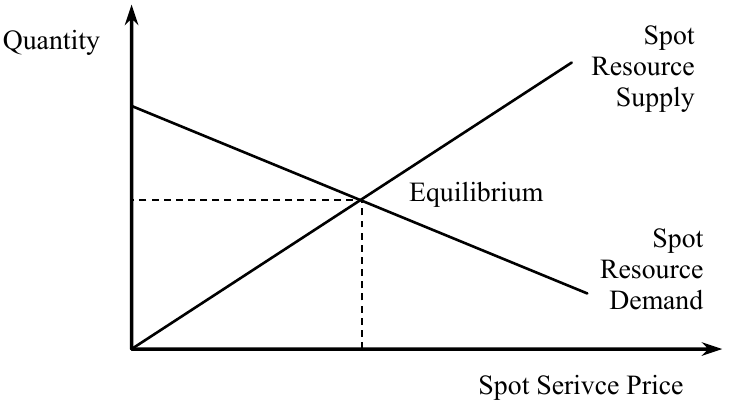}
\caption{Cloud spot pricing from the perspective of economics.}
\label{fig_economicModel}
\end{figure}

\begin{figure}
  \centering
  \subfloat[Electronical input/output view.]{
    \label{fig:subfigEngineering:io} 
    \includegraphics[width=3.5cm]{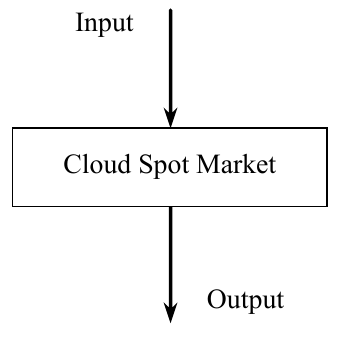}}
 \qquad
  \subfloat[Mechanical state space view.]{
    \label{fig:subfigEngineering:ss} 
    \includegraphics[width=3.8cm]{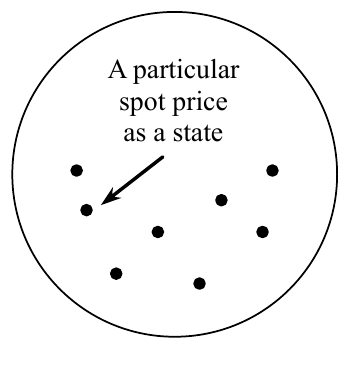}}
  \caption{Cloud spot pricing from the perspective of engineering.}
  \label{fig:subfigEngineering} 
\end{figure}

When it comes to the research method, two opposite approaches can be distinguished in the relevant studies: the black-box investigations are usually based on the statistical analyses of Amazon's spot price traces, while the white-box investigations are mostly based on theoretical models of market participant activities. In particular, behind the explanatory theories, we find two popular classes of domain-specific techniques employed from the economics and engineering disciplines respectively. From the perspective of economics, Cloud spot pricing has been treated as various auctions and games to reach some equilibrium (e.g., the Prisoner Dilemma game \citeS{Di_Valerio_2013}), as shown in Fig.~\ref{fig_economicModel}. From the perspective of engineering, researchers have modeled Cloud spot pricing as electronical dynamics with input/ouput factors (e.g., $k$ parallel $M/M/1$ queues \citeS{Abhishek_Kash_2012}) or mechanical dynamics within a state space (e.g., a faulty machine \citeS{Song_Zafer_2012}), as illustrated in Fig.~\ref{fig:subfigEngineering}.

After all, it is clear that there is no one-size-fits-all theory for revealing the secrets of Cloud spot pricing. As a result, we find that different types of theories could supplement each other for understanding the de facto pricing mechanism. For example, as a descriptive theory, the likelihood of employing reserve prices was claimed to be evidence against Amazon's market-driven mechanism \citeS{Ben-Yehuda_2013}; nevertheless, considering the investment cost mentioned by some explanatory theories, setting a reserve price would be reasonable for auctioneers to prevent losses when demand was very low \citeS{Zaman_Grosu_2013}. However, we admit that none of the economic models can explain the similar price fluctuations of Amazon's different spot resources. 

Within the same theory type, we find that it is hard to aggregate the collected data due to the diverse and sometimes contradictory statements. Yet a positive side effect of the contradictory theories is that people can use them to cross assess different primary studies. For example, since the systematic predictability analysis proved that normal distribution was inappropriate for spot price approximation \citeS{Zhao_Pan_2012}, the assumption of normally distributed prices could have flawed the corresponding work on spot service bidding \citeS{Mazzucco_Dumas_2011}.

In particular, given the rational and intelligent participants in the Cloud market, it has been identified that economic models would be more suitable than conventional models in the context of Cloud spot pricing \citeX{Prasad_Rao_2014}. On the other hand, the engineering models could suffer from simplified while improper assumptions. For example, the study \citeS{Abhishek_Kash_2012} assumed that there were always jobs coming up, and its simulation suggested that a well-chosen fixed price would beat spot prices in terms of revenue maximization. In practice, however, there could be jobs attracted by spot prices only, and for these jobs a ``well-chosen" fixed price would not be easily chosen well in advance. Thus, the conclusions in the study \citeS{Abhishek_Kash_2012} might have been flawed by its unsuitable assumptions, which confirms our previous discussions in Section \ref{subsubsec:RQ1_discusion}.

\subsection{Techniques to address limitations of Cloud spot pricing (RQ3)}
\label{subsec:RQ3results}
Given the consensus on the major drawbacks of spot pricing for Cloud consumers, most of the relevant studies attempted to address the compromised availability of spot services. We have identified six typical fault-tolerance techniques developed from the consumers' perspective, and two improvement approaches proposed from the provider's point of view. Although some papers report composite or integrated fault-tolerance solutions (e.g., a combination of four OS-level mechanisms is proposed in \citeS{He_Shenoy_2015}), we focus on the individual techniques only, as briefly described in the following subsections.

\subsubsection{Checkpointing}
Checkpointing seems to be the most popular fault-tolerance technique to boost overall computing performance and productivity \citeS{Jung_Lim_2014}, and sometimes to save the cost \citeS{Poola_Ramamohanarao_2014}, when employing spot resources. In brief, the checkpointing technique allows consumers to save their intermediate work when spot service is terminated. A set of practical checkpointing strategies developed by the relevant studies are:

\begin{itemize}
    \item	\textit{Adaptive Scheme:} Checkpoints are taken or skipped at a 10-minute frequency based on the estimation of expected recovery time in case of future service interruptions \citeS{Yi_Andrzejak_2012}.
    \item	\textit{Application-centric Scheme:} This scheme is based on a sophisticated event generation system that delivers decision to either take checkpoints or terminate spot instances \citeS{Khatua_Mukherjee_2013_paper}. 
    \item	\textit{Enhanced Adaptive Incremental Scheme:} This is a revised version of the adaptive scheme based on an adjusted Markov model that takes into account both service revocations and hardware failures \citeS{Jangjaimon_Tzeng_2013}.
    \item	\textit{Hourly Scheme:} Checkpoints are taken periodically prior to the beginning of each spot service hour \citeS{Voorsluys_Buyya_2012,Yi_Andrzejak_2012}.
    \item	\textit{Rising edge-driven Scheme:} Checkpoints are taken after every increase (rising edge) of spot prices even if the out-of-bid event does not happen \citeS{Leslie_Lee_2013,Yi_Andrzejak_2012}.
\end{itemize}

\subsubsection{Duplication or redundancy}
It has been identified that Cloud spot service may only be suitable for short-term jobs due to the possible frequent interruptions \citeS{Chaisiri_Kaewpuang_2011}. To increase the chance of satisfying longer jobs' deadline constraints, the duplication technique creates one replica of each job that could run for more than one hour, and the replicas are supposed to be deployed with different instance type/datacenter combinations \citeS{Voorsluys_Buyya_2012}. Considering that each zone has a strong dependency on its own price history, the study \citeS{Marathe_Harris_2014} particularly suggested the across-zone redundancy as a complementary fault-tolerance mechanism.

\subsubsection{Lineage-based recovery}
Lineage-based recovery works by re-executing the computation of the failed nodes \citeS{Dadashov_Cetintemel_2014}. Compared to the other fault-tolerance techniques, lineage-based recovery generally incurs little overhead during normal execution. However, it does not work if many spot instances fail at the
same time. Therefore, this technique suggests employing different instance types to reduce the risk that all spot instances fail simultaneously.

\subsubsection{Migration}
The migration technique can be viewed as an improvement of checkpointing. By using the checkpointing technique, the suspended jobs have to wait until the spot service is resumed. To avoid waiting for re-acquiring the same spot resources, the migration technique suggests rebidding at a comparable per-core price for different types of instances even from a different datacenter \citeS{Jung_Chin_2013,Yi_Andrzejak_2012}. If successful, the whole image of a spot instance will be migrated from its original physical host to another host that runs virtual machines at other prices and even with other pricing schemes \citeS{Chen_Lee_2014}.

\subsubsection{Nested virtualization}
Nested virtualization indicates the architecture where a virtual machine runs inside another virtual machine and a job runs inside the nested virtual machine \citeS{He_Shenoy_2015}. In essence, the purpose of nested virtualization is to facilitate the aforementioned migration, by having the complete control over the nested virtual machine without requiring any privileged access to the native one.

\subsubsection{Stop-and-redo model}
Unlike the lineage-based recovery technique, the stop-and-redo model will restart a job from scratch even if only one spot instance fails, which is usually for a traditional database system deployed in spot markets \citeS{Dadashov_Cetintemel_2014}. Thus, the best application strategy driven by the stop-and-redo model is to use as many spot instances as necessary to finish the job in less than one hour.

\subsubsection{Service scaling down}
Recall that the termination of spot instances would not only interrupt the consumers' applications but also reduce the provider's profit (cf.~Section \ref{subsubsec:providerLimit}). From the provider's perspective, the study \citeS{Dawoud_Takouna_2012} suggested scaling down a spot service proportionally to the increase in its price, so as to improve the spot service's availability. In other words, when necessary to free some compute resources, the provider may sacrifice some capacity of its service for not terminating the relevant spot instances.

\subsubsection{Hybrid spot instance}
To relieve the overhead burden of not charging the last partial hour, the study \citeS{Sadashiv_Kumar_2014} proposed a Hybrid Spot Instance approach that could share the fault tolerance cost between the service provider and consumers, and meanwhile could make the spot service more reliable. The Hybrid Spot Instance approach is essentially based on the aforementioned checkpointing technique, and it allows stretching the user bid till checkpointing is done when an out-of-bid situation occurs. 

\subsubsection{Discussion}
As mentioned previously, unexpected service interruption is the most significant concern for consumers in the spot market. Then, fault-tolerance techniques would be crucial to help increase practitioners' confidence in using Cloud spot services. In fact, although the widely accepted use cases of spot service are small-scale, flexible and delay-tolerant jobs, suitable fault-tolerance techniques have been demonstrated to be able to keep the spot service's availability level as high as using on-demand instances \citeS{Guo_Chen_2015,Ostermann_Prodan_2012}, and to make spot instances possible to support highly-reliable always-on online applications \citeS{Guo_Chen_2015,He_Shenoy_2015}.

However, the existing fault-tolerance techniques are not free \citeS{Sadashiv_Kumar_2014}. For example, checkpointing would inevitably occupy extra fixed-price resource (e.g., storage), while duplication and migration would require more compute resources. Moreover, a rough trend we have found is: the more sophisticated the approach is, the more cost it may incur. Therefore, the target of service availability should be well balanced, otherwise in the worst case the fault-tolerance cost could overwhelm the economic benefits from employing Cloud spot service. 

Ideally, it is also possible to ask the provider like Amazon to modify the spot market and improve its spot service's reliability \citeS{Marathe_Harris_2014}, for example, using the hybrid spot mechanism to allow finalizing a checkpoint before any termination. From the provider's perspective, unfortunately, such market modifications would not be desirable if they lead to fewer consumers entering the usually more profitable on-demand market. 

\section{Threats to validity}
\label{sec:validity_threats}
Although we strove to perform the review activities as rigorously and objectively as possible, the findings of this SLR study might still have been affected by certain limitations, as listed below. Readers may need to consider these validity threats when applying the reported findings to their own work.

\subsection{Completeness}
It is possible that our paper selection does not exhaustively cover all the relevant studies. First of all, as revealed in Section \ref{subsubsec:RQ1_discusion}, different researchers could investigate Cloud spot pricing from different angles with various terms and concepts. As a result, the studies that have irregular descriptions of their empirical investigations might have been missed out. Second, to balance the possible workload with coverage, we searched the five most popular electronic libraries instead of looking up every possible literature resource. To alleviate these two issues, we employed a manual search by snowballing the references to further identify as many missing relevant studies as possible. 

Last but not the least, in order to maintain a reasonable effort on literature search, we performed automated search through the metadata (namely titles, keywords and abstracts) rather than full text of the papers. However, it has been identified that the search precision would be reduced dramatically by scanning full text \cite{Dieste_Griman_2009}, and the automated search could inevitably miss relevant studies due to the limitations of search engines \cite{Brereton_Kitchenham_2007}. Therefore, we consider this as a known observation instead of a limitation of this SLR study.

\subsection{Reviewers reliability}
Recall that our work concentrates on the empirical evidence of spot pricing in the Cloud computing domain only (cf.~Section \ref{subsec:research_scope}). When collecting and synthesizing evidence, for this comparative investigation, we tried to distinguish the statements about spot pricing in the Cloud-specific sense from those in the generic sense. Thus, different reviewers involved in this study could possess slightly different opinions in some cases. 

To reduce the possible bias, we cross-reviewed the collected data, and further discussed unsure issues in our group meetings. Since this study is an international collaboration with team members separated geographically, we largely resorted to Skype and telephone for our group meetings. Although the modern communication facilitates are convenient, the different time zones still made it difficult for our discussion on a frequent basis, which might also incur possible bias in the review process. Compared to our previous practices, we put more effort on organizing meetings and making our discussions more efficient in this study. To the best of our knowledge, few SLR studies have reported their experiences of remote collaboration on conducting SLR. Such a challenge could bring new research opportunities in our EBSE community. 

\subsection{Evidence strength}
When assessing the quality of the relevant studies (cf.~Section \ref{subsec:quality_assessment}), we supposed that their empirical work like simulations and data analyses were valid, and the corresponding statements and findings were trustable. Nevertheless, it is possible that the primary studies could have been flawed by improper assumptions or unsuitable modeling of the Cloud spot market. This could also be the reason why there are contradictory statements about Cloud spot pricing. For example, the conclusion drawn by study \citeS{Abhishek_Kash_2012} about the spot pricing scheme is contrary against the other studies. 

The ideal approach to judging the evidence strength would be to replicate and compare those empirical investigations. However, the replication verification is out of the scope of this paper. Thus, we highlight this possible limitation only as a reminder for readers to take into account evidence strength when interpreting the results of this SLR study.

\subsection{Industrial gap}
As previously mentioned, there are few Cloud providers using the spot pricing scheme to offer service in industry, and the only one provider Amazon might have employed artificial algorithm instead of real market-driven mechanism to fluctuate spot prices \citeS{Ben-Yehuda_2013}. Such a phenomenon could bring bias in the existing studies and enlarge the gap between academic outcomes and industrial needs. For example, when proposing descriptive and predictive theories (cf.~Section \ref{subsubsec:descriptive} and \ref{subsubsec:predictive}), nearly all the relevant studies took or had to take Amazon's price trace to perform empirical invistigations. As for the explanatory and prescriptive theories (cf.~Section \ref{subsubsec:explanatory} and \ref{subsubsec:prescriptive}), there is little evidence to show that those sophisticated pricing system design and assumptions are practically aligned with the real market \citeS{Kushwaha_Simmhan_2014}.

We have carefully balanced our research scope to reduce the impact of this threat. On the one hand, to avoid discussions sticking to Amazon, we tried to distinguish the research focus between on general pricing mechanism and on Amazon's own policies (e.g., Section \ref{subsubsec:consumerlimit}). On the other hand, to make our work closer to the current situation in industry, we have excluded studies on precurement auction (or reverse auction) that considers competition among Cloud providers (cf.~Section \ref{subsec:research_scope}). 

In addition, we believe that such an industrial gap also reminds our community of potential research opportunities in this domain. For example, the industry still desires a widely-accepted and easily-implementable pricing mechanism for the Cloud spot market. 

\section{Conclusions and future work}
\label{sec:conclusion}
Appropriate pricing schemes and techniques are crucial for developing and maintaining a successful and sustainable Cloud ecosystem. Through collecting, assessing and analyzing the relevant evidence from 61 primary studies, our systematic literature review (SLR) shows that the academic community strongly advocates the emerging Cloud spot market. Most of the relevant studies reported encouraging discussions about and/or analyses of applying spot pricing to Cloud computing. Although the complexity in backend mechanisms could be a long-term obstacle for both offering and employing Cloud spot services, a set of fault-tolerance techniques have been developed to help reduce some limitations from the consumers' side. In the review, only one study favored fixed pricing from the perspective of Cloud providers according to its queuing theory-based analysis and simulation \citeS{Abhishek_Kash_2012}. However, we have doubts about such work not only because of the opposite majority opinion, but also because its improper assumptions could have weakened the validity of the simulation result. 

Meanwhile, this SLR also reveals that there is still a lack of practical market-driven mechanisms to support Cloud spot pricing. A possible dilemma is: economic models would be able to better reflect the Cloud spot market, while the corresponding optimization problems could be NP-hard or NP-complete. As for the de facto spot market with respect to Amazon, although reserve price is reasonable from the economic angle, there is little evidence clearing the doubt about Amazon's spot-price scheme. By synthesizing four types of relevant theories, we confirm the existence of a large gap between the sophisticated pricing models imposed by the existing studies and the real needs in reality, because both providers and consumers would prefer simple mechanisms in practice.

Overall, the findings of this SLR have raised our confidence in the research topic of Cloud spot pricing. We would also expect this report to be able to increase practitioners' confidence in joining the Cloud spot market. Our future work will then be unfolded in two directions: first, we will focus on the reusability of the extracted data, e.g., developing a factor checklist for cost-benefit analysis of being a spot pricing player; second, we will put more effort on investigating practical and easily deployable spot mechanisms.




\appendix

\section{Quality assessment scores of the primary studies}
\label{sec:quality_scores}
The quality assessment has generated a set of detailed scores for each primary study, as listed in Table \ref{tbl>quality_assessment}.

\begin{table*}[!t]\footnotesize
\renewcommand{\arraystretch}{1.3}
\centering
\caption{\label{tbl>quality_assessment}Detailed score card for the quality assessment of the 61 primary studies.}
\begin{tabular*}{.85\textwidth}{ @{\extracolsep{\fill}} lcccccccccc}
\hline

\hline
\multirow{2}{*}{\textbf{Paper}} & \multicolumn{5}{c}{\textbf{Quality of the Study}} & \multirow{2}{*}{\textbf{Sum (1-5)}} & \multicolumn{3}{c}{\textbf{Strength of the Evidence}} & \multirow{2}{*}{\textbf{Sum (6-8)}}\\
\cline{2-6}
\cline{8-10}
 & \textbf{QA1} & \textbf{QA2} & \textbf{QA3} & \textbf{QA4} & \textbf{QA5} & & \textbf{QA6} & \textbf{QA7} & \textbf{QA8} & \\
\hline
\citeS{Abhishek_Kash_2012}& 1 & 1 & 1 & 1 & 0.5 & 4.5 & 1 & 1 & 1 & 3\\
\citeS{Andrzejak_Kondo_2010} & 1 & 0.5 & 1 & 1 & 1 & 4.5 & 1 & 0 & 0 & 1 \\ 
\citeS{Abundo_Valerio_2014} & 0.5 & 1 & 1 & 1 & 1 & 4.5 & 1 & 0.5 & 0 & 1.5\\
\citeS{Abundo_Valerio_2015}  & 1 & 1 & 0.5 & 1 & 1 & 4.5 & 1 & 0 & 0 & 1 \\
\citeS{Binnig_Salama_2015} & 1 & 0.5 & 0.5 & 1 & 1 & 4 & 1 & 0.5 & 0 & 1.5 \\
\citeS{Ben-Yehuda_2013} & 1 & 1 & 1 & 1 & 1 & 5 & 1 & 1 & 0.5 & 2.5\\ 
\citeS{Chohan_Castillo_2010} & 1 & 0 & 1 & 1 & 1 & 4 & 0.5 & 0.5 & 0 & 1\\
\citeS{Chaisiri_Kaewpuang_2011} & 0.5 & 1 & 0.5 & 1 & 1 & 4 & 1 & 0.5 & 0.5 & 2\\ 
\citeS{Chen_Lee_2014} & 1 & 1 & 1 & 1 & 1 & 5 & 1 & 1 & 1 & 3 \\
\citeS{Chichin_Vo_2014} & 1 & 1 & 0.5 & 1 & 1 & 4.5 & 1 & 0 & 0 & 1 \\
\citeS{Dadashov_Cetintemel_2014} & 1 & 0 & 0 & 0 & 0 & 1 & 1 & 0.5 & 0 & 1.5 \\
\citeS{Di_Valerio_2013}& 1 & 0.5 & 1 & 1 & 1 & 4.5 & 1 & 1 & 0 & 2\\

\citeS{Dawoud_Takouna_2012} & 1 & 1 & 1 & 1 & 1 & 5 & 0.5 & 0.5 & 0 & 1\\ 
\citeS{Guo_Chen_2015} & 1 & 1 & 0.5 & 1 & 1 & 4.5 & 1 & 1 & 1 & 3\\
\citeS{He_Shenoy_2015} & 1 & 0.5 & 0.5 & 1 & 1 & 4 & 1 & 0.5 & 0 & 1.5\\
\citeS{Jung_Chin_2011}  & 1 & 1 & 1 & 1 & 1 & 5 & 1 & 0.5 & 0 & 1.5\\
\citeS{Jung_Chin_2013}  & 1 & 1 & 1 & 1 & 1 & 5 & 0.5 & 0.5 & 0 & 1\\
\citeS{Jung_Lim_2014} & 1 & 1 & 1 & 1 & 0.5 & 4.5 & 1 & 0.5 & 0 & 1.5\\
\citeS{Jangjaimon_Tzeng_2013} & 1 & 1 & 1 & 1 & 1 & 5 & 1 & 0.5 & 0 & 1.5\\
\citeS{Javadi_Thulasiram_2013}& 1 & 1 & 1 & 1 & 1 & 5 & 0.5 & 0 & 0 & 0.5\\ 
\citeS{Kantere_Dash_2011} & 0.5 & 1 & 0.5 & 1 & 1 & 4 & 1 & 1 & 1 & 3\\
\citeS{Karakus_Li_2014} & 0.5 & 0.5 & 0 & 1 & 1 & 3.5 & 1 & 1 & 1 & 3\\
\citeS{Khatua_Mukherjee_2013_paper} & 1 & 1 & 1 & 1 & 1 & 5 & 1 & 0.5 & 0 & 1.5\\ 
\citeS{Karunakaran_Sundarraj_2013} & 1 & 0 & 1 & 1 & 1 & 4 & 0.5 & 0 & 0 & 0.5\\ 
\citeS{Kushwaha_Simmhan_2014} & 0.5 & 1 & 0.5 & 1 & 1 & 4 & 1 & 1 & 1 & 3 \\
\citeS{Kaminski_Szufel_2015} & 1 & 0.5 & 0.5 & 1 & 1 & 4 & 1 & 0 & 0 & 1 \\
\citeS{Karunakaran_Sundarraj_2015} & 1 & 0 & 1 & 1 & 1 & 4 & 1 & 0.5 & 0 & 1.5\\
\citeS{Liu_2011} & 1 & 0.5 & 1 & 1 & 1 & 4.5 & 1 & 0 & 0 & 1\\
\citeS{Leslie_Lee_2013}& 1 & 1 & 1 & 1 & 1 & 5 & 0.5 & 0.5 & 0.5 & 1.5\\
\citeS{Lim_Thakur_2014} & 1 & 1 & 0.5 & 1 & 1 & 4.5 & 1 & 0.5 & 0 & 1.5\\
\citeS{Mazzucco_Dumas_2011}  & 1 & 1 & 1 & 1 & 1 & 5 & 1 & 0.5 & 0.5 & 2\\ 
\citeS{Marathe_Harris_2014} & 1 & 1 & 1 & 1 & 1 & 5 & 1 & 0.5 & 0 & 1.5\\
\citeS{Menache_Shamir_2014} & 1 & 1 & 1 & 1 & 1 & 5 & 1 & 1 & 0 & 2 \\
\citeS{Mattess_Vecchiola_2010} & 1 & 1 & 1 & 1 & 1 & 5 & 0.5 & 0.5 & 0.5 & 1.5\\
\citeS{Ostermann_Prodan_2012} & 1 & 1 & 0.5 & 1 & 1 & 4.5 & 0.5 & 0.5 & 0.5 & 1.5\\
\citeS{Poola_Ramamohanarao_2014} &  0.5 & 0 & 1 & 1 & 1 & 3.5 & 1 & 1 & 1 & 3 \\
\citeS{Qin_Wu_2012} & 1 & 0.5 & 0.5 & 1 & 1 & 4 & 1 & 1 & 1 & 3\\
\citeS{Ribas_Furtado_2015} & 0.5 & 0.5 & 1 & 1 & 1 & 4 & 1 & 1 & 1 & 3 \\
\citeS{Singh_Dutta_2015} & 1 & 0 & 0.5 & 1 & 1 & 3.5 & 1 & 1 & 0 & 2\\
\citeS{Sharma_Lee_2015}  & 1 & 1 & 1 & 1 & 1 & 5 & 1 & 0 & 0 & 1 \\
\citeS{Sadashiv_Kumar_2014}  & 1 & 0 & 0 & 1 & 1 & 3 & 1 & 0 & 0 & 1\\
\citeS{Sowmya_Sundarraj_2012} & 1 & 0.5 & 1 & 1 & 1 & 4.5 & 1 & 1 & 0.5 & 2.5\\
\citeS{Song_Yao_2013}& 1 & 1 & 1 & 1 & 1 & 5 & 0.5 & 0.5 & 0 & 1\\ 
\citeS{Song_Zafer_2012} & 1 & 1 & 1 & 1 & 1 & 5 & 1 & 1 & 0 & 2\\ 
\citeS{Shi_Zhang_2014} & 0.5 & 1 & 1 & 0.5 & 1 & 4 & 1 & 1 & 0 & 2\\
\citeS{Taifi_2012} & 1 & 1 & 1 & 1 & 1 & 5 & 1 & 0.5 & 0 & 1.5\\
\citeS{Taifi_Shi_2011} & 1 & 1 & 1 & 1 & 1 & 5 & 1 & 0 & 0 & 1\\
\citeS{Tang_Yuan_2012} & 1 & 1 & 1 & 1 & 1 & 5 & 0.5 & 0 & 0 & 0.5\\
\citeS{Voorsluys_Buyya_2012}& 1 & 1 & 1 & 1 & 1 & 5 & 0.5 & 0.5 & 0 & 1\\
\citeS{Voorsluys_Garg_2011} & 1 & 1 & 1 & 1 & 1 & 5 & 0.5 & 0.5 & 0 & 1\\

\multicolumn{11}{r}{\small\sl continued on next page}\\
\end{tabular*}
\end{table*}

\begin{table*}[!t]\footnotesize
\renewcommand{\arraystretch}{1.3}
\centering
\begin{tabular*}{.85\textwidth}{ @{\extracolsep{\fill}} lcccccccccc}
\multicolumn{11}{l}{\small\sl Table \ref{tbl>quality_assessment} continued from previous page}\\
\hline

\hline
\multirow{2}{*}{\textbf{Paper}} & \multicolumn{5}{c}{\textbf{Quality of the Study}} & \multirow{2}{*}{\textbf{Sum (1-5)}} & \multicolumn{3}{c}{\textbf{Strength of the Evidence}} & \multirow{2}{*}{\textbf{Sum (6-8)}}\\
\cline{2-6}
\cline{8-10}
 & \textbf{QA1} & \textbf{QA2} & \textbf{QA3} & \textbf{QA4} & \textbf{QA5} & & \textbf{QA6} & \textbf{QA7} & \textbf{QA8} & \\
\hline
\citeS{Wee_2011}  & 1 & 0.5 & 1 & 1 & 1 & 4.5 & 0.5 & 0.5 & 0.5 & 1.5\\
\citeS{Wang_Qi_2013}& 1 & 1 & 0.5 & 1 & 1 & 4.5 & 0.5 & 0 & 0 & 0.5\\
\citeS{Wang_Ren_2012} &  1 & 1 & 1 & 1 & 1 & 5 & 0.5 & 0.5 & 0 & 1\\
\citeS{Wallace_Turchenko_2013} & 1 & 0 & 1 & 1 & 1 & 4 & 0.5 & 0 & 0 & 0.5\\
\citeS{Xu_Li_2013} & 1 & 1 & 0.5 & 1 & 1 & 4.5 & 1 & 1 & 0 & 2\\
\citeS{Yi_Andrzejak_2012} &  1 & 1 & 1 & 1 & 1 & 5 & 1 & 0.5 & 0 & 1.5\\
\citeS{Zaman_Grosu_2011}& 1 & 1 & 0.5 & 1 & 1 & 4.5 & 1 & 1 & 0 & 2\\
\citeS{Zaman_Grosu_2013} & 1 & 1 & 1 & 1 & 1 & 5 & 1 & 1 & 1 & 3\\
\citeS{Zhang_Gurses_2011} & 1 & 1 & 1 & 0.5 & 1 & 4.5 & 1 & 1 & 0 & 2\\
\citeS{Zhang_Li_2014} & 1 & 1 & 0.5 & 1 & 1 & 4.5 & 1 & 1 & 1 & 3 \\
\citeS{Zhao_Pan_2012} &  1 & 1 & 1 & 1 & 1 & 5 & 0.5 & 0.5 & 0 & 1\\

\hline
\textbf{Total} & 57 & 47.5 & 49.5 & 59 & 59 & 273 & 52.5 & 34.5 & 15 & 102\\

\textbf{Average} & 0.93 & 0.78 & 0.81 & 0.97 & 0.97 & 4.47 & 0.86 & 0.57 & 0.25 & 1.67\\
\hline

\hline
\end{tabular*}
\end{table*}

\section{Selected primary studies}
\begingroup
\renewcommand{\section}[2]{}
\bibliographystyleS{alpha}
\bibliographyS{SLR_Selected_Studies}
\endgroup

\section{Explanation of the typical excluded studies}
\label{Appendix>Explanation_Exclusion}

\begin{table*}[!t]\footnotesize
\renewcommand{\arraystretch}{1.35}
\centering
\caption{\label{tbl>3aa}Explanation of several typical excluded studies}
\begin{tabular}{l l p{2.2cm}}
\hline

\hline
\textbf{Paper} & \textbf{Brief Explanation} & \textbf{Corresponding Exclusion Criteria}\\
\hline
\citeX{Chun_Choi_2013} & This paper analyzes two types of fixed pricing schemes. & (5)\\
\citeX{Silva_Neto_2012} & This is a secondary study on accounting models for Cloud computing. & (8)\\
\citeX{Karunakaran_Krishnaswamy_2015} & This paper performs a survey type of work on Cloud computing from the business view. & (8)\\

\citeX{Khatua_Mukherjee_2013} & This is a poster presentation without clear specification about validation. & (9)\\

\citeX{Meinl_Anandasivam_2010} & This paper emphasizes the imperfect competition among multiple Cloud providers. & (2)\\

\citeX{Prasad_Rao_2014} & This paper investigates reverse auction that involves multiple providers. & (2)\\
\citeX{Polverini_Ren_2013} & This paper focuses on the energy consuming when discussing pricing for Cloud computing. & (1)\\

\citeX{Patel_Sarje_2012} & This paper studies load balancing of resource requirements in a federated Cloud environment. & (2)\\

\citeX{Sharma_Thulasiram_2012} & This paper discusses Cloud resource pricing in a generic sense. & (6)\\

\citeX{Tian_Wang_2012} & This is a short paper without clear specification about validation. & (9)\\

\citeX{Vanmechelen_Depoorter_2011} & This paper investigates the pricing mechanism for Grid computing systems. & (4)\\

\citeX{Zaman_Grosu_2010} & This is an old-version duplication of the selected study \citeS{Zaman_Grosu_2013}. & (7)\\

\citeX{Zaman_Grosu_2012} & This paper focuses on the mechanism implementation without clearly discussing spot pricing. & (3)\\

\citeX{Zhao_Li_2014} & This paper considers job pricing instead of Cloud service pricing.  & (4)\\

\citeX{Zhou_Sun_2015} & This paper does not specify any benefit or limitation of Cloud spot pricing.  & (3)\\ 
\hline

\hline
\end{tabular}
\end{table*}

See Table \ref{tbl>3aa}. To save space, we only show several typical excluded publications. Most excluded papers were discussed in our group meetings. Note that the studies listed here may also be used as a clue for readers to further identify useful information.

\section{Typical excluded primary studies}
\begingroup
\renewcommand{\section}[2]{}
\bibliographystyleX{alpha}
\bibliographyX{SLR_Excluded_Studies}
\endgroup

\bibliographystyle{model1b-num-names}
\setbiblabelwidth{99}
\bibliography{SLR_Ref}

\end{document}